\documentclass[17pt]{article}

\usepackage{graphicx}
\usepackage{natbib}
\usepackage{authblk}
\usepackage{amsthm}

\usepackage[colorinlistoftodos]{todonotes}
\usepackage{stackrel}

\usepackage{gensymb}
\usepackage{algorithm}
\usepackage{algpseudocode}

\usepackage{amssymb,amsfonts,amsmath}

\newcommand{\bR}{{\mathbb R}}
\newcommand{\bD}{{B}}

\newcommand{\cR}{{\cal R}}

\newcommand{\bbf} {{\boldsymbol f }}

\newcommand{\by}{{\boldsymbol y}}
\newcommand{\bw}{{\boldsymbol w}}
\newcommand{\prob}{\mathbb{P}}
\newcommand{\mE}{{\mathbb{E}}}

\newtheorem{thm}{Theorem}[section]

\begin{document}

\title{Threshold Selection for Total Variation Denoising}


\author[1]{Sylvain Sardy and H. Monajemi
\thanks{S. Sardy is with the Department of Mathematics, Universit\'{e} de Gen\`{e}ve, Gen\`{e}ve, Switzerland,
        e-mail: sylvain.sardy@unige.ch\\
        H. Monajemi is with the Department of Statistics, Stanford University, California, USA,
        email: monajemi@stanford.edu}}


\maketitle


\begin{abstract}
Total variation (TV) denoising is a nonparametric smoothing method that has good properties for preserving sharp edges and  contours in objects with spatial structures like natural images.
The estimate is sparse in the sense that TV reconstruction leads to a piecewise constant function with a small number of jumps.
A threshold parameter controls the number of jumps and the quality of the estimation. In practice, this threshold is often selected by minimizing
a goodness-of-fit criterion like cross-validation,
which can be costly as it requires solving the high-dimensional and non-differentiable TV optimization problem many times.
We propose instead a two step adaptive procedure via a connection to large deviation of stochastic processes.
We also give conditions under which TV denoising achieves exact segmentation.
We then apply our procedure to denoise a collection of 1D and 2D test signals verifying the effectiveness of our approach in practice.     
\end{abstract}


\section{Introduction}

Consider a function  $f$ defined on a $d$-dimensional lattice of $M=N_1 \times \ldots \times N_d$ points (e.g., a square image for $d=2$ and $N_1=N_2$).
Let $f_i$ denote the value of $f$ at location $i \in \{1,\dots, M\}$. Suppose we are given data 
\begin{equation}\label{eq:obs_model}
 {y_i}= f_i + { \epsilon_i} \quad {\rm with} \quad \epsilon_i\stackrel{{\rm i.i.d.\ }}{\sim} {\rm N}(0,\sigma^2), \quad  i=1,\dots, M,
\end{equation}
where the noise ${\epsilon_i}$ is independent and identically distributed (i.i.d.)\ Gaussian. 
The denoising problem is to recover the unknown vector $\bbf=(f_1, \dots, f_M)$ from noisy observations $\by=(y_1, \dots y_M)$.
A variety of denoising methods exists, such as smoothing splines \citep{Duchon76,Wahb:spli:1990}, kernel estimators \citep{Muller87},
Markov random field \citep{GG84,Besag86}, wavelets \citep{Dono94b}. 
More recent image denoising methods include patch-based methods \citep{KB06}, learned dictionaries \citep{EA06}, BM3D \citep{DFKE07},
global image denoising \citep{TM14}, topological denoising \citep{GJRSSW14}, sub-Riemannian anisotropic smoothing \cite{Miolane15,Duits2016} and references therein.

 

This article mainly concerns total variation (TV) denoising \citep{ROF92}, which is closely linked to Laplace Markov random field \citep{SardyTseng04}. 
Originally, \citet{ROF92} proposed (isotropic) TV for image denoising and observed its
ability in preserving sharp edges without introducing much spurious oscillations.
They proposed a gradient-projection algorithm for finding the solution to TV minimization by solving a time-dependent partial differential equation.
TV is still of interest for image denoising;
for instance, \citet{chambolle04,chambolle05} proposed a faster algorithm based on a dual formulation for both isotropic and anisotropic total variation.
 \citet{Harc:Lvy-:mult:2010} and \citet{TVChambolleDuvalPeyrePoon16} studied the segmentation properties of TV.

TV for image denoising can be defined as follows. Consider an image $\bbf$ and a pixel $f_{i,j}$
with its ``east'' pixel $f_{i+1,j}$ and ``north'' pixel $f_{i,j+1}$.
Then, for a given penalty $\lambda \geq 0$,
the discretized version of isotropic TV defines the estimate $\hat \bbf_\lambda$ as the solution to
\begin{eqnarray*}
\min_{{\bbf}\in \mathbb{R}^M} && \frac{1}{2}\| {\bf y} - {\bbf} \|_2^2\\
&&+ \lambda \sum_{i=1}^{N_1-1} \sum_{j=1}^{N_2-1} \sqrt{ (f_{i,j}-f_{i+1,j})^2 + (f_{i,j}-f_{i,j+1})^2},
\end{eqnarray*}
while the \emph{anisotropic} version solves
$$
\min_{{\bbf}\in \mathbb{R}^M} \frac{1}{2}\| {\bf y} - {\bbf} \|_2^2+ \lambda \sum_{i=1}^{N_1-1} \sum_{j=1}^{N_2-1}  |f_{i,j}-f_{i+1,j}| + |f_{i,j}-f_{i,j+1}|.
$$
The former is reminiscent of group lasso \citep{Yuan:Lin:mode:2006} and the latter of lasso \citep{Tibs:regr:1996}. In other words
the isotropic version sets both directional gradients to zero at once, while the anisotropic version can detect an edge in one direction
and a flat region in another direction.
Since anisotropic TV allows different smoothness on the horizontal and vertical directions, 
we favor anisotropic TV.
More generally, for any $d\geq 1$ dimensional function, anisotropic TV  solves the following optimization problem:  
\begin{equation} \label{eq:TVprimal}
\min_{{\bbf} \in \mathbb{R}^M }  \frac{1}{2} \| \by - \bbf \|_2^2  + \lambda   \| \bD \bbf \|_1,
\end{equation}
where  $\bD^T=[\bD_1^T, \bD_2^T, \dots, \bD_d^T]$ is the sparse derivative matrix of size $M(d-\sum_{i=1}^d 1/N_i)\times M$
that is constructed by concatenating directional finite difference matrices $B_i$ for $i=1,\dots,d$. Each row of $\bD$ thus contains two nonzero entries $+1$ and $-1$.


Although the optimization problem~(\ref{eq:TVprimal}) is strictly convex, the nondifferentiability  and high dimensionality of the TV cost function make it difficult to solve. A number of ideas that include Bregman iterations, second-order cone programming, primal-dual interior point methods, dual formulation, proximal operators and alternating direction method of multipliers have been proposed in the literature.
See~\citet{Goldstein09,Goldfarb05,Hintermuller06,SardyTseng04,chambolle04,Wahlberg12,FISTA09,micchelli10}, and the references cited therein.  

A less settled problem is the selection of the penalty parameter $\lambda$ that has a thresholding effect: the larger $\lambda$ the more entries in $B\bbf$ are set to zero.
The quality of reconstruction clearly depends on the choice of the threshold $\lambda$.
\citet{Mamm:van:loca:1997} derived asymptotic properties such as rates of convergence in bounded variation function classes for an ideal $\lambda$.
\citet{DJMontanari:2013} provide the threshold $\lambda$ corresponding to the minimax risk of TV denoising in the context of compressed sensing phase transitions,
where the coordinatewise risk (or mean squared error) 
of the TV estimate $\hat{\bbf}_\lambda$ with respect to the true function $\bbf$ is defined by
\begin{equation}\label{eq:MSE}
\cR_{\lambda} =\frac{ \mE \|  \bbf - \hat{\bbf}_\lambda\|_2^2 }{M}.
\end{equation}
In practice, given a finite number of observations and no oracle information, one needs to deduce $\lambda$ from the data.
Traditional approaches calculate a goodness-of-fit criterion for several $\lambda$'s and select the best one.

These include cross validation, empirical Bayes \citep{SardyTseng04}, and Stein unbiased risk estimation (SURE) \citep{Stein:1981}.
\citet{TibshTaylor:2011} showed that the degrees of freedom for TV is the number of connected components ${\rm NCC}(\hat \bbf_\lambda)$ of
the estimated solution~$\hat \bbf_\lambda$ leading to
\begin{equation} \label{eq:SURENCC}
{\rm SURE}(\lambda)= M^{-1} \|\by - \hat \bbf_\lambda \|_2^2+2\sigma^2 M^{-1} {\rm NCC}(\hat \bbf_\lambda) -  \sigma^2.
\end{equation}
These approaches can be costly for large problem sizes because they require solving the optimization problem~\eqref{eq:TVprimal} for many $\lambda$.
Moreover, calculating the number of connected components ${\rm NCC}(\hat \bbf_\lambda)$ precisely can impose a challenge because the numerical solution to~\eqref{eq:TVprimal} is only approximate, and so determining the number of connected components requires appropriate rounding to significant figures. The reader is referred to the reproducible code (see section \ref{rr}) distributed with this article for the approximations we used to determine NCC in our numerical simulations.

We propose instead a procedure that requires solving the optimization problem~(\ref{eq:TVprimal}) only twice.
In Section~\ref{sct:dual}, we define the universal threshold for TV.
We study the univariate case (i.e., $d=1$) in Section~\ref{sct:d=1}. In particular, we derive the universal threshold in Section~\ref{subsec:constant-function} and the two-step adaptive universal threshold in Section~\ref{subsct:TVsmooth}.
In Section~\ref{subsct:exactseg} we investigate whether TV can achieve exact segmentation  for piecewise constant functions.
Section~\ref{subsct:ex1D} presents numerical results to study the performance of our threshold selection procedure in comparison with other selection methods and to corroborate our theoretical results on exact segmentation.
In Section~\ref{sct:d>=2} we extend the idea of an adaptive universal threshold
to dimensions $d \geq 2$. In particular, we report new stochastic processes that are more challenging to study analytically because they are not well understood by the currently available large deviation theories. Nonetheless, in Section~\ref{subsct:multi} we establish numerical procedures that give appropriate choice of $\lambda$ for multi-dimensional functions.
Section~\ref{subsct:imageex} quantifies the performance of our two-step procedure to denoise images.
Finally, conclusions and future research suggestions are presented in section~\ref{sct:conclusion}. The research is reproducible (see Section~\ref{rr}). Proofs are postponed to appendices.


%
%

\section{Dual formulation and fitting a constant}
\label{sct:dual}

In this section, we consider data that are generated by adding noise to a constant function. Our aim is to establish a connection between the threshold $\lambda$ and the probability that the TV estimate~$\hat \bbf_\lambda$ be the best constant fit to the data.
In practice, the underlying function to recover is rarely constant; however, understanding this connection guides us---%
as we will see later in this article---in selecting appropriate thresholds for recovering piecewise constant functions and other heterogeneous functions that can be well approximated by a piecewise constant one. 

The role of $\lambda$ is better understood by considering 
the Lagrangian dual problem for~(\ref{eq:TVprimal}):
\begin{equation}  \label{eq:TVdual}
\min_{\bbf, \bw}  \frac{1}{2} \| \bbf\|_2^2  \quad {\rm s.t.} \quad \left \{
\begin{array}{l}
 \by-\bD^T \bw=\bbf \\
 \| \bw \|_{\infty} \le \lambda
\end{array}
\right . .
\end{equation}
See \citet{SardyTseng04} or \cite{chambolle04} for a derivation of the dual problem. The solution to the TV denoising problem can thus be seen as
$
\hat{\bbf}_\lambda  = \by - \hat{\boldsymbol \epsilon} ,
$
where the noise $\hat{\boldsymbol \epsilon}$ is modeled as $B^T \bw$ for some
$\bw=(\bw^1, \bw^2, \dots \bw^d) \in \bR^{M(d-\sum_{i=1}^d 1/N_i)}$ satisfying  $\| \bw \|_{\infty} \le \lambda$.
Since the constant vector spans the kernel of $\bD$, solving~\eqref{eq:TVdual} with the constraint that $\bbf$ is a constant vector amounts to solving
$\by-B^T \bw=\bar y {\bf 1}$, where $\bar y$ is the mean of the data. The smallest threshold~$\lambda$ allowing a solution of this type is thus the smallest $\| \bw \|_{\infty}$ for all $\bw$ satisfying $\by-B^T \bw=\bar y {\bf 1}$.


For stochastic input $\by$ centered around a constant vector, the dual variable $\bw$ that satisfies $\by-B^T \bw=\bar y {\bf 1}$ and has smallest Sup-norm defines a stochastic process.
Consequently if one chooses the TV threshold $\lambda$ to be larger than $\| \bw \|_{\infty}$ with a probability tending to one with the size of the data,
then the TV estimate is provably the best parametric fit $\hat \bbf_\lambda=\bar y {\bf 1}$, producing no jumps with high probability.
This is analogous to the approach adopted in \cite{Dono94b} for selecting the threshold in wavelet denoising, which has asymptotic minimax properties \citep{Dono95asym}.
In what follows we study the empirical processes defined by the TV dual solution $\bw$, and we will see how their large deviation properties can help us select appropriate thresholds for TV denoising problems. We are now ready to define the \emph{universal threshold} for TV denoising.

%

\bigskip
{\bf Definition 1}~(Universal Threshold): Consider the random variable
\begin{equation} \label{eq:lambdaNd>1}
  \Lambda=  \min_{\bw} \| \bw \|_\infty \quad {\rm s.t.} \quad {\bf Y}- B^{\rm T} \bw = \bar Y {\bf 1},
\end{equation}
where ${\bf Y}$ is given by our model~(\ref{eq:obs_model}) under the assumption that $\bbf$ is a constant function on a $d$-dimensional lattice
of dimension $M=N_1 \times \ldots \times N_d$, and $\bar Y={\rm 1}^{\rm T} {\bf Y}/M$ is the mean of the data.
The  \emph{universal threshold} $\lambda_{M}$ for TV is the $(1-\alpha_M)$-quantile of  $\Lambda$ for some small $\alpha_M$.
By analogy with the universal threshold for wavelet smoothing \citep{Dono94b}, we choose $\alpha_M=O(1/\sqrt{\log P_M})$, where $P_M=M(d-\sum_{i=1}^d 1/N_i)$ is TV's degrees of freedom;
that is, the number of finite differences involved in TV's penalty or the number of rows in $B$.

\bigskip

\bigskip
{\bf Property 1.} Suppose that the underlying function to recover is constant and equal to $\mu$ on the $d$-dimensional lattice, that is $Y_n\stackrel{\rm i.i.d.\ }\sim{\rm N}(\mu, \sigma^2)$; then the TV estimate $\hat \bbf_{\lambda_M}$ calculated with the universal threshold $\lambda_M$ defined in Definition~1 has the property that
$$
{\mathbb P}( \hat \bbf_{\lambda_M}=\bar Y {\bf 1})=1-\alpha_M.
$$

\bigskip

We derive a closed form expression for $\lambda_M$ in the univariate case in Section~\ref{sct:d=1} and provide an approximate expression
in higher dimensions in Section~\ref{sct:d>=2}. 
Note that the universal threshold depends on the noise level $\sigma$. When necessary we estimate $\sigma$ with the median absolute deviation of the finite differences re-scaled to be consistent
under Gaussian white noise as originally proposed by \citet{Dono94b}. Namely, 
\begin{equation} \label{eq:MADsigma}
 \hat \sigma= 1.4826/\sqrt{2} \cdot {\rm median}_j(|(B \by)_j - {\rm median}_j (B \by)_j|).
\end{equation}

\section{Univariate denoising ($\lowercase{d}=1$)}
\label{sct:d=1}

\subsection{Universal threshold for constant function}\label{subsec:constant-function}


In dimension~$d=1$, the matrix $B=B_1$ of finite differences has more columns ($N$) than rows ($N-1$).
Note that this is no longer true in higher dimensions (see Section~\ref{sct:d>=2}).
The kernel of $B$ is spanned by the constant vector.
Consequently $B^T$ has full column rank and the linear equation
${\bf Y}-B^{\rm T} \bw = \bar Y {\bf 1}$ in \eqref{eq:lambdaNd>1} has a unique solution 
$\bw= (BB^T)^{-1}B{\bf Y}$.
 \citet{SardyTseng04} observed that random process $\bw/\sqrt{N}$ has the distribution of a discrete Brownian bridge when data ${\bf Y}$ are centered around the same constant 
 (or equivalently, the underlying function is constant over the lattice).
So $\Lambda/\sqrt{N}=\| \bw \|_\infty/\sqrt{N}$ satisfies
\begin{align*}
{\mathbb P}(\Lambda/\sqrt{N} \leq u)&\geq & {\mathbb P}(\| {\mathbb U}\|_\infty\leq u)
\\
&=&1-2\sum_{k=1}^\infty (-1)^{k+1} \exp(-2k^2u^2)
\\
&\geq&  1-2 \exp(-2u^2),
\end{align*}
where ${\mathbb U}$ is the Brownian bridge. This inequality leads to the following closed form expression for the universal threshold.

\bigskip
{\bf Property 2.} In dimension $d=1$, the universal threshold for TV is $\lambda_N=\frac{\sigma}{2} \sqrt{N \log \log N}$.
Under the assumption the data are white noise added to a constant function,
the TV estimate produces $\hat \bbf_{\lambda_N}=\bar y {\bf 1}$
with probability at least $1-\alpha_N$ with $\alpha_N=2/\sqrt{\log N}$. 

\subsection{Adaptive universal threshold for piecewise constant function}
\label{subsct:TVsmooth}


The function to recover is rarely the constant function, but many functions can be well approximated by a piecewise constant function.


So suppose the function $f$ sampled in \eqref{eq:obs_model} is piecewise constant with $L$ constant pieces according to the following definition.

\bigskip
{\bf Definition 2.} An $L$-piecewise constant vector $\bbf^{\rm pc}$ has entries defined by
\begin{eqnarray} 
f^{\rm pc}_i&=&h_l  \quad {\rm for} \quad  N_{\bullet (l-1)}<i\leq N_{\bullet l}, \\ \label{eq:f=h}
 &  &  \ l=1, \ldots, L, \ i=1,\ldots,N, \nonumber
\end{eqnarray}
where $N_k$ is the number of observations in the $k$th constant piece, $N_{\bullet l}=\sum_{k=1}^l N_k$ denotes the location of the $l$-th jump for $l=1,\ldots,L-1$, with $N_{\bullet 0}=0$ and $N_{\bullet L}=N$.
The average number of samples per level is $\bar N_L=N/L$,
and the jump signs are $s_j={\rm sign}(h_{j+1}-h_{j})$  with the convention that $s_0=s_L=0$.

\bigskip

Note that $L$ corresponds to the number of connected components mentioned in~\eqref{eq:SURENCC} of an $L$-piecewise constant vector $\bbf^{\rm pc}$, 
that is $L={\rm NCC}(\bbf^{\rm pc})$.

When the problem size gets large and the function to be estimated is piecewise constant with a fixed number $L$ of levels,
the leading terms of the TV cost function~(\ref{eq:TVprimal})
are the sum of the squared residuals within each of the $L$ levels, along with their corresponding finite differences summed in absolute value.
So a proxy to the TV cost function in that case is to consider $L$ independent smoothing terms.
The threshold $\lambda_N=\frac{\sigma}{2}\sqrt{N\log \log N}$ derived for $N$ data can be used to fit each term by adapting the formula
to the average number of sample $\bar N_L=N/L$ data per level. This approach leads to the definition below.

\bigskip
{\bf Definition 3.} In dimension $d=1$, the  \emph{adaptive universal threshold} for TV is $\lambda_{N,L}=\frac{\sigma}{2} \sqrt{\bar{N}_L\log \log \bar{N}_L}$,
where $L$ is the number of levels of the underlying function to recover, $\bar N_L=N/L$, and $N$ is the problem size.

\bigskip

To get an estimate of $L$ in practice, one could use
$$
\hat L_{\bf y}= \sum_{i=1}^{N-1} {\bf 1}(|(B { \bf y})_i|>\sigma \sqrt{2} z_{1-0.025/(N-1)} ),
$$
where $\sqrt{2}$ stems from the $+1$ and $-1$ in each row of $B$ and $1-0.025/(N-1)$ is a Bonferroni correction at significance level $\alpha=0.05$.
Though this estimate of $L$ performs well for large jumps, it is essentially powerless when there exists true jumps of small magnitude (for example, in smooth functions), whence underestimating the number of correct jumps.  
Another possibility is to rely on the screening property of lasso with $\lambda=\lambda_N$ \citep{BuhlGeer11}  
to provide a set of estimated jumps that may include the true ones with high probability. Namely, we would count the number of jumps $\hat L_0$ in $\hat \bbf_{\lambda_N}$ according to
$$
\hat L_0= \sum_{i=1}^{N-1} {\bf 1}(|(B\hat \bbf_{\lambda_N})_i|\neq 0).
$$
This tends to overestimate the true number of jumps because of the spurious jumps detected by the lasso estimator, as also corroborated in the next section.
A better choice is to estimate the number of jumps by 
\begin{equation} \label{eq:hatL}
 \hat L= \sum_{i=1}^{N-1} {\bf 1}(|(B\hat \bbf_{\lambda_N})_i|>\sigma \sqrt{2/N} z_{1-0.025/(N-1)}).
\end{equation}
This formula calibrates the variance of the estimate on that of the average, namely $\sigma^2/N$; although conservative, this approach allows to get rid of many spurious jumps. 
We find that $\hat L$ provides a good approximation for the true number of jumps $L^0$ when the function is piecewise constant, 
and that typically $\hat L_{\bf y}\leq L^0\leq  \hat L \leq \hat L_0$.

To summarize, the two-step procedure solves~\eqref{eq:TVprimal} with the universal threshold $\lambda=\lambda_N$ to get an estimate $\hat L$ of $L$,
then solves TV with the adaptive threshold $\lambda_{N,\hat L}$.
We investigate in Section~\ref{subsct:ex1D} the risk performance of TV with the adaptive universal threshold to estimate not only piecewise constant functions,
but also smoother ones.



\subsection{A threshold for segmentation with TV?}
\label{subsct:exactseg}

The problem of segmentation consists in finding regions where approximation of the underlying function $f$ by a constant is reasonable. A standard approach consists in optimizing the combinatorial problem of probing all possible jump locations and minimizing the corresponding least squares fits \citep{CIS-59789,CIS-89410,CIS-166846}.
Motivated by the  piecewise constant nature of the TV estimate, \citet{Harc:Lvy-:mult:2010} studied the segmentation properties of TV. Among interesting results
regarding the distance of the estimated jump locations to the true ones, they claim that perfect estimation of the change points cannot happen. We contend this result is not true for all piecewise constant functions.

Indeed, we show below
that exact segmentation happens with high probability for an explicit threshold we provide, when the underlying function is piecewise constant as in Definition~2,
provided adjacent jumps alternate sign. 

The KKT conditions for TV to achieve exact segmentation are the following.



\begin{thm} \label{thm:exact-segmentation-KKT}
Assume model~(\ref{eq:obs_model}) for a piecewise constant function according to Definition~2.
The KKT conditions for the TV estimate $\hat {\bbf }=(\hat {\bbf}_1, \ldots, \hat {\bbf}_L)$ to do exact segmentation are
\begin{equation} \label{eq:KKT}
\left \{
\begin{array}{rcl}
\hat {\bbf}_l&=&\hat h_l {\bf 1}_{N_l}, \quad l=1,\ldots,L \\
\hat h_l &=& \bar y_l + (s_l-s_{l-1}) \frac{\lambda}{N_l}, \quad l=1,\ldots,L \\
s_j &=& {\rm sign}(\hat h_{j+1}-\hat h_{j}), \quad j=1,\ldots,L-1\\
\bw &\in& [-\lambda,\lambda]
\end{array}
\right . ,  
\end{equation}
where $s_0=s_L=0$ by convention and the dual variables $\bw=(w_1,\dots,w_{N-1})$ are given by
\begin{align*}
w_{N_{\bullet (l-1)}+i}=- \sum_{k=1}^i  y_{N_{\bullet (l-1})+k} +i \hat h_l  + \lambda s_{l-1},\\
i=1,\ldots,N_l, \ \ l=1,\ldots,L.& & 
\end{align*}
\end{thm}
\bigskip

Theorem \ref{thm:exact-segmentation-KKT} can be used to determine whether there exists $\lambda$ to achieve exact segmentation.
If exact segmentation is feasible, then one may hope to derive an empirical threshold that achieves exact segmentation with high probability.
We have already shown in Section~\ref{subsec:constant-function} that when $L=1$ (i.e., the constant function), 
choosing the \emph{universal threshold} of Property~2
leads to $\hat {\bbf}_{\lambda_N}=\bar y {\bf 1}$ with high probability.  


It is interesting to study whether exact segmentation is still feasible 
when the underlying function 
is made of $L>1$ constant pieces. 
The following theorem states a quite surprising result that a necessary condition for exact TV segmentation is an alternating sign condition.

\bigskip


\begin{thm}\label{thm:conditionES}
Assume model~(\ref{eq:obs_model}) for a piecewise constant function according to Definition~2.
A necessary condition for the  TV estimate to achieve exact segmentation (ES) is that the unknown piecewise constant function
alternates jump signs, that is $s_{l+1}=-s_l$ for $l=1,\ldots,L-2$.
If moreover the jumps are high enough, then
choosing $\lambda_N^{{\rm ES}}=\sigma N_{\max} \Phi^{-1}(1-\alpha_N/2)$ (where $\Phi$ is the Gaussian distribution function) with $N_{\max}=\max_{l=1,\ldots,L} N_l$
achieves it with probability at least $\pi_0^{ES} = (1-2\alpha_N)^{L-2} (1-\alpha_N)^2$ for any $\alpha_N \in [0,1/2)$.
\end{thm}
\bigskip

Considering the bias of the TV estimate with $\lambda_N^{{\rm ES}}$ reveals that the jumps $h_l$ must be high enough with respect to the noise level $\sigma$
to achieve exact segmentation. To see this, consider the situation where each piecewise constant level is equally sampled
(that is, $N_1=\ldots=N_L=N/L$) and where the alternating jumps sign condition holds (that is, $|s_l-s_{l-1}|=2$).
By the KKT conditions~(\ref{eq:KKT}), the bias of $\hat h_l$ is $(s_l-s_{l-1}) \frac{\lambda_N^{\rm ES}}{N_{\max}}=2 \sigma  \Phi^{-1}(1-\alpha_N/2)$ for $l=1,\ldots,L$.
Consequently, if the jump is not high enough with respect to the threshold, that is, if the condition 
\begin{equation}  \label{eq:hstar}
h_l> h^* \quad {\rm with} \quad  h^*= 4 \sigma  \Phi^{-1}(1-\alpha_N/2)  
\end{equation}
is not satisfied (there are one downward and one upward biases, hence a factor 4), then exact segmentation will be achieved with low probability.
These new results on the limitation of TV to perform exact segmentation are linked to
the \emph{irrepresentable condition} derived for lasso with a general regression matrix \citep{Zhao:2006:MSC:1248547.1248637}.
Recently \citet{TVChambolleDuvalPeyrePoon16} studied the segmentation property of 2D isotropic TV and made a precise mathematical account of the regions where the sharp edge preserving property of TV works, as opposed to regions where spurious staircasing effects may occur.

In practice the alternate jump sign condition is unrealistic for univariate signals. Take the {\tt blocks} function for instance \citep{Dono94b};
this piecewise constant function has alternate jump signs except in one instance in the center of its domain (see right plot of Figure~\ref{fig:testFunc}), which prevents exact segmentation.
But this does no prevent the use of TV to provide an approximate segmentation.

\subsection{Examples} \label{subsct:ex1D}

\begin{figure*}[ht]
\centering
\includegraphics[width=4.5in,height=2.5in,angle=0]{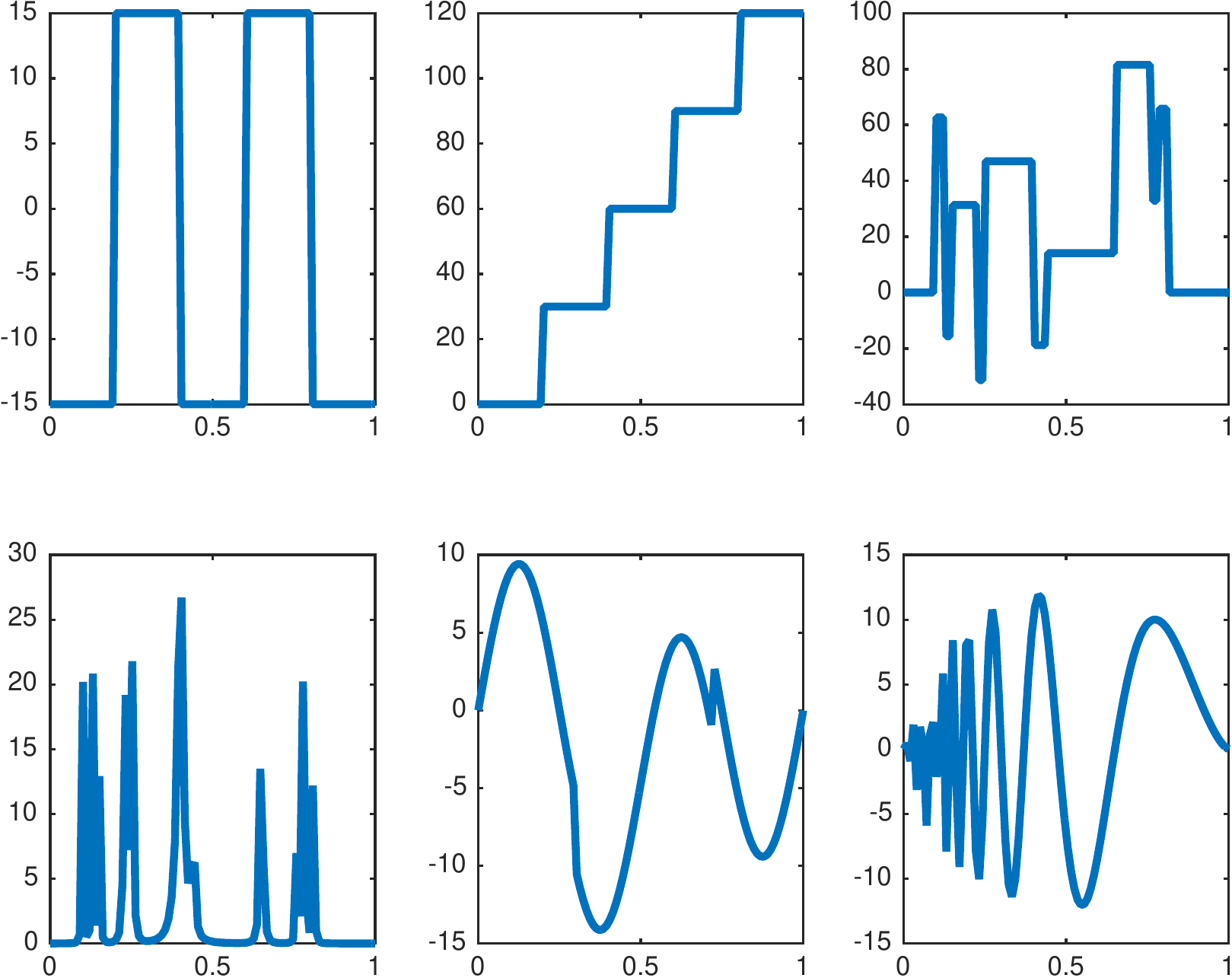}
\caption{One-dimensional test functions used in the simulations.\label{fig:testFunc}}
\end{figure*}

We study the empirical performance of the universal threshold, the adaptive universal threshold and the exact segmentation threshold. In our experiments described below, we simulate data with independent Gaussian noise added to six different test functions
four of which originally appeared in \citet{Dono94b}. These test functions, depicted in Figure \ref{fig:testFunc}, are {\tt blocks}$(N,\text{SNR})$,  {\tt bumps}$(N,\text{SNR})$, {\tt heavisine}$(N,\text{SNR})$ and {\tt Doppler}$(N,\text{SNR})$ for denoising, and {\tt battlements}($N,L,h$) and {\tt staircases}$(N,L,h)$ for segmentation. Here, $L$ is the number of constant pieces, $h$ is the height of jumps, $N$ is the length of the equispaced sampling grid, and $\text{SNR}$ is the signal to noise ratio.
We consider two types of experiments:  
\begin{itemize}
\item \textit{Denoising.} We estimate the mean squared error~\eqref{eq:MSE}  for 
problem sizes $N\in \{10^2, 10^3,10^4 \}$ with respective number of Monte Carlo runs $M\in\{500,50,5\}$, for the four test functions  {\tt blocks}, {\tt bumps}, {\tt heavisine} and {\tt Doppler}
with SNR=7, as well as the {\tt zero}-constant function. Four selection rules for the threshold $\lambda$ are compared: oracle (optimal $\lambda$ in terms of $\ell_2$-loss), Stein unbiased risk estimation,
empirical Bayes and the adaptive threshold
defined in Section~\ref{subsct:TVsmooth}. To compare the methods fairly, the true $\sigma$ is provided.

The results of the simulation summarized in Table~\ref{table-1DMSE} show that the adaptive threshold compares favorably with empirical Bayes, and is sometimes even better than minimizing the Stein unbiased risk estimate.
As expected the proposed method works remarkably well for a piecewise constant function like {\tt blocks} or {\tt zero}.
The adaptive threshold (which requires solving TV with only two values of $\lambda$) also has the advantage of being computationally more efficient than minimizing a criterion.

\begin{table*}[ht]
\centering
\caption{Coordinatewise risk \eqref{eq:MSE} (*100) results of Monte Carlo simulation for denoising univariate fonctions.} 
\label{table-1DMSE}
\begin{tabular}{l|rrrr}
  \hline
  & Oracle & SURE & EB &  Adaptive $\lambda_{N, \hat L}$ \\
 \hline
{\bf blocks} &  &  &  &  \\ 
  $N=10^2$ & 38.7 & 42.0 & 51.6 & 42.3 \\ 
  $N=10^3$ & 6.5 & 6.8 & 13.2 & 6.6 \\ 
  $N=10^4$ & 0.8 & 0.9 & 4.4 & 0.8 \\ 
  {\bf bumps} &  &  &  &  \\ 
  $N=10^2$ & 70.4 & 73.2 & 74.5 & 103.1 \\ 
  $N=10^3$ & 36.0 & 37.3 & 37.7 & 36.5 \\ 
  $N=10^4$ & 10.7 & 10.9 & 10.8 & 12.0 \\ 
  {\bf heavisine} &  &  &  &  \\ 
  $N=10^2$ & 54.4 & 57.0 & 58.1 & 63.0 \\ 
  $N=10^3$ & 12.4 & 12.7 & 13.0 & 13.7 \\ 
  $N=10^4$ & 2.6 & 2.7 & 3.0 & 3.2 \\ 
  {\bf Doppler} &  &  &  &  \\ 
  $N=10^2$ & 80.6 & 82.7 & 81.7 & 85.7 \\ 
  $N=10^3$ & 34.9 & 36.2 & 39.1 & 35.1 \\ 
  $N=10^4$ & 7.9 & 8.1 & 7.9 & 8.9 \\ 
  {\bf zero} &  &  &  &  \\ 
  $N=10^2$ & 1.5 & 2.2 & 24.1 & 1.5 \\ 
  $N=10^3$ & 0.1 & 0.3 & 4.9 & 0.1 \\ 
  $N=10^4$ & 0.0 & 0.0 & 2.3 & 0.0 \\ 
   \hline
\end{tabular}
\end{table*}
 
%

\item \textit{Exact segmentation and screening.} 
We consider three piecewise constant test functions: {\tt battlement}, {\tt staircase} and {\tt blocks} plotted  on Figure~\ref{fig:testFunc}.
The jump height is chosen to be $2h^*$ with $h^*$ defined in~\eqref{eq:hstar} for {\tt battlement} and {\tt staircase};
likewise {\tt blocks} is rescaled so that the smallest jump height is $2h^*$.
With such height, Theorem \ref{thm:conditionES} predicts segmentation occurs with high probability for {\tt battlement} only because it is the only function among the three test functions that satisfies the alternating jumps sign condition.
We also consider two other heights for {\tt battlement}. Namely, $h^*$ for which the theory suggests that the exact segmentation is less likely compared to $2h^*$, and $h^*/10$ for which exact segmentation is unlikely.

To verify these results empirically, 
we estimate the probability of exact segmentation 
\begin{align*}
 \pi^{ES}(\lambda|f,N,L) =&\text{$\prob$\{the set of jump locations}& \\
&\text{\ \ of $\hat \bbf_\lambda$ matches those of $\bbf$\} },&
\end{align*}
and of screening
\begin{align*}
 \pi^{S}(\lambda|f,N,L) =&\text{$\prob$\{the set of jump locations}& \\
&\text{\ \ of $\hat \bbf_\lambda$ include those of $\bbf$\} }.&
\end{align*}

for each test function, problem size $N$ and number of constant pieces $L$.
These probabilities depend on the threshold selected:
we consider the threshold $\lambda_N^{{\rm ES}}$ of Theorem~\ref{thm:conditionES} and the universal threshold $\lambda_N$ of Definition~2. 

\begin{table*}[ht]
\scriptsize
\centering
\caption{Monte Carlo simulation for exact segmentation and screening with TV using $\lambda_N^{ES}$ (first $5$ columns)  and $\lambda_N$ (last $3$ columns). Columns 1: theoretical lower bound $\pi_0^{ES}$ of Theorem~3.2 (when exact segmentation is possible). Column 2 and 6: probability of exact segmentation. Column 3 and 7: probability of screening. Column 4: true number of levels in piecewise constant function $f$. Column 5 and 8: expected number of levels in $\hat f$.} 
\label{table-1DSegmentation}
\begin{tabular}{l||ccccc|ccc}
  \hline
  &\multicolumn{5}{c|}{$\lambda_N^{ES}$ for exact segmentation ($\alpha_N=0.05$)} & \multicolumn{3}{c}{$\lambda_N$ for screening} \\
 & $\pi_0^{ES}$ &  $\hat{\pi}^{ES}$ &  $\hat{\pi}^S$ &  $L^0$ & ${\mathbb E}(\hat L)$ &  $\hat{\pi}^{ES}$ &  $\hat{\pi}^S$ & ${\mathbb E}(\hat L)$  \\ \hline  \hline
{\bf battlements(N,5,$\boldsymbol{2h^*}$)} &  &  &  &  &  &  &  &  \\ 
  $N=10^2$ & 0.66 & 0.95 & 1 & 5 & 5 & 0 & 1 & 7 \\ 
  $N=10^3$ & 0.66 & 0.99 & 1 & 5 & 5 & 0 & 1 & 11 \\ 
  $N=10^4$ & 0.66 & 0.95 & 1 & 5 & 8 & 0 & 1 & 19 \\ 
  {\bf battlements(N,5,$\boldsymbol{h^*}$)} &  &  &  &  &  &  &  &  \\ 
  $N=10^2$ & 0.66 & 0.93 & 1 & 5 & 5 & 0 & 1 & 7 \\ 
  $N=10^3$ & 0.66 & 0.94 & 1 & 5 & 5 & 0 & 1 & 11 \\ 
  $N=10^4$ & 0.66 & 0.92 & 1 & 5 & 6 & 0 & 1 & 19 \\ 
  {\bf battlements(N,5,$\boldsymbol{h^*/10}$)} &  &  &  &  &  &  &  &  \\ 
  $N=10^2$ &  & 0 & 0 & 5 & 1 & 0 & 0.2 & 3 \\ 
  $N=10^3$ &  & 0 & 0 & 5 & 1 & 0 & 0.2 & 9 \\ 
  $N=10^4$ &  & 0 & 0 & 5 & 1 & 0 & 0.3 & 16 \\ 
  {\bf staircase(N,5,$\boldsymbol{2h^*}$)} &  &  &  &  &  &  &  &  \\ 
  $N=10^2$ &  & 0 & 1 & 5 & 8 & 0 & 1 & 8 \\ 
  $N=10^3$ &  & 0 & 1 & 5 & 11 & 0 & 1 & 13 \\ 
  $N=10^4$ &  & 0 & 1 & 5 & 18 & 0 & 1 & 18 \\ 
  {\bf blocks(N,$\equiv\boldsymbol{2h^*}$)} &  &  &  &  &  &  &  &  \\ 
  $N=10^2$ &  & 0 & 0 & 12 & 11 & 0 & 1 & 14 \\ 
  $N=10^3$ &  & 0 & 0 & 12 & 12 & 0 & 1 & 20 \\ 
  $N=10^4$ &  & 0 & 0 & 12 & 16 & 0 & 1 & 53 \\ 
   \hline
\end{tabular}
\normalsize
\end{table*}

As established by Theorem~3.2, Table~\ref{table-1DSegmentation}
for {\tt battlements}(N,5,$2h^*$) shows that exact segmentation occurs with probability higher than $\pi_0^{\rm ES}$ (defined in \eqref{eq:pi0ES}, here with $\alpha_N=0.05$) when three conditions hold:
the alternate jump sign condition holds,
the size of the jumps is  significantly larger than $h^*$ (here $2h^*$) and the threshold is chosen to $\lambda_N^{\rm ES}$.
As expected the result no longer holds when the height of the jumps is much lower, like  $h^*/10$.
We also show with the {\tt staircase} and {\tt blocks} functions that when the alternate jump sign condition fails, 
the threshold $\lambda_N^{{\rm ES}}$  fails to provide TV denoising with the exact segmentation property.
As far as screening is concerned, the threshold $\lambda_N$ guarantees it with probability one for the high signal to noise ratio considered here.
The expected number $\hat L$ defined in~\eqref{eq:hatL} of estimated steps in the underlying function is also reported in Table~\ref{table-1DSegmentation}.
In the right columns, we see that screening is guaranteed with the universal threshold $\lambda_N$ but that spurious jumps are detected because ${\mathbb E}(\hat L)>L^0$,
the true number of jumps. 
\end{itemize}

\section{Multivariate denoising ($\lowercase{d}>1$)}
\label{sct:d>=2}

\subsection{Multivariate universal threshold for TV}
\label{subsct:multi}

The good denoising property of the adaptive universal threshold in dimension $d=1$ reported in Table~\ref{table-1DMSE} calls for its extension to higher dimensions. 
All that is needed to adapt the two step TV estimate to denoise images ($d=2$) and data on higher dimensional lattices is the distribution of $\Lambda$ in~\eqref{eq:lambdaNd>1}.
Once the distribution of $\Lambda$ is known, then taking the appropriate upper quantiles provides the universal thresholds and the adaptive universal thresholds.


To the best of our knowledge, existing probability results do not provide an expression for the distribution of $\Lambda$ when $d\geq 2$, however.
The difficulty mainly stems from the fact that there exists an infinite number of dual vectors $\bw$ satisfying  ${\bf Y}-B^{\rm T} {\bw}= \bar Y {\bf 1}$ because $B^{\rm T}$ has more columns than rows when the dimension $d$ of the lattice is larger 
or equal to two.
And minimizing the sup norm among all solutions ${\bw}$ in Definition~1 makes the problem also hard.

In the following, we investigate the \emph{empirical} distribution of $\Lambda$ and derive the empirical universal threshold on $d$-dimensional square lattices for $d=2$ and $d=3$,
for a range of problem sizes $N$.
To do so, we rewrite the optimization problem~\eqref{eq:lambdaNd>1} as a linear programming problem:
$$
\min_{\bw, \lambda}\  ({\bf 0}^{\rm T}, 1) \left ( \begin{array}{c} \bw \\ \lambda \end{array} \right ) \quad {\rm s.t.} \quad \left \{
\begin{array}{l}
{\bf y}-\bar y {\bf 1} \leq  B^{\rm T} \bw \leq {\bf y}-\bar y {\bf 1} \\
\bw - \lambda {\bf 1} \leq {\bf 0} \\
\bw + \lambda {\bf 1} \geq {\bf 0} \\
\lambda \geq 0
\end{array} \right . .
$$
Importantly the minimum to~\eqref{eq:lambdaNd>1} in $\lambda$ is uniquely defined (but may be reached for several $\bw$). 
We use the optimization solver {\tt MOSEK} for solving this problem in {\tt MATLAB}. 

\begin{table*}[!ht]
\centering
\caption{$p$-values of the Gumbel versus GEV likelihood ratio test} 
\label{table-LRtest}
\begin{tabular}{|c|llllllll|}
  \hline
  & \multicolumn{8}{c|}{Problem Size $N$} \\
 \hline dimension &8 &16 &32 &64 &128 &256 &512 &1024 \\
 \hline
2 & 0.286 & 0.546 & 0.097 & 0.035 & 0.291 & 0.903 & 0.549 & 0.430 \\ 
  3 & 0.811 & 0.349 & 0.483 & 0.033 &  &  &  &  \\ 
   \hline
\end{tabular}
\end{table*}

\begin{figure*}[ht]
\centering
$
\begin{array}{llll}
\includegraphics[width=1.05in]{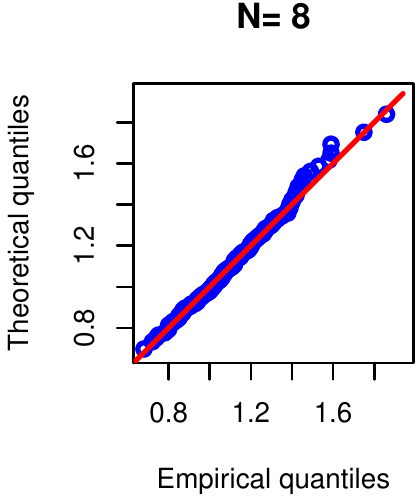}&
\includegraphics[width=1.05in]{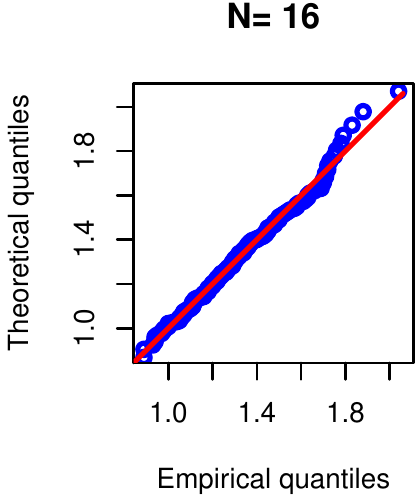}&
\includegraphics[width=1.05in]{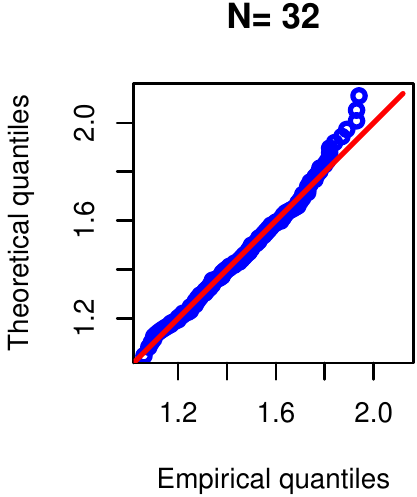}&
\includegraphics[width=1.05in]{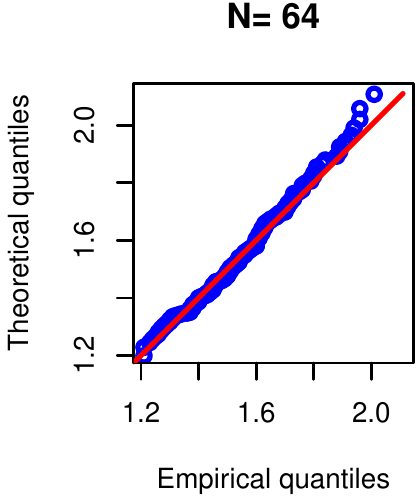}\\
\includegraphics[width=1.05in]{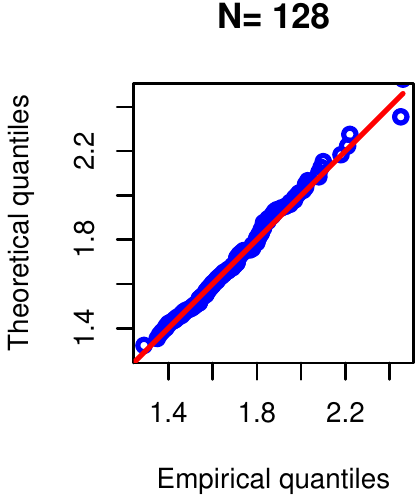}&
\includegraphics[width=1.05in]{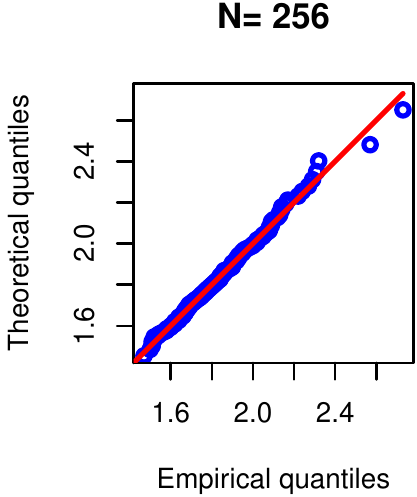}&
\includegraphics[width=1.05in]{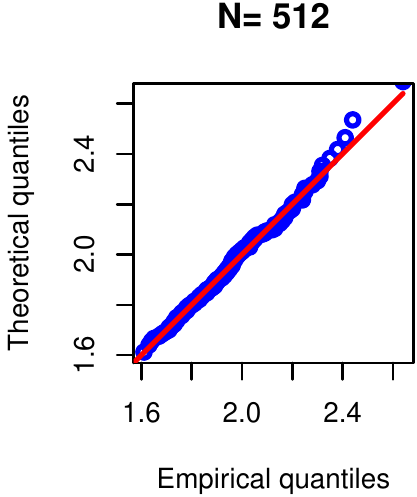}&
\includegraphics[width=1.05in]{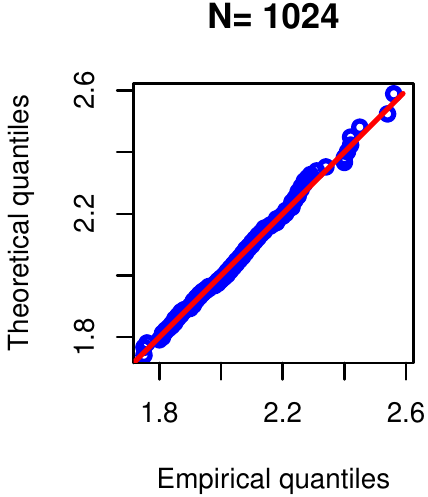}
\end{array}
$
\caption{Quantile-quantile plot of empirical distribution of $\Lambda$ versus the fitted Gumbel distribution for 2D problems of sizes $N\in\{8, 16, 32, 64, 128, 256, 512, 1024\}$. In each panel, the red line indicates the $y=x$ identity line.}
\label{fig:qq-plot-2D}
\end{figure*}

\begin{figure*}[ht]
\centering
$
\begin{array}{llll}
\includegraphics[width=1.05in]{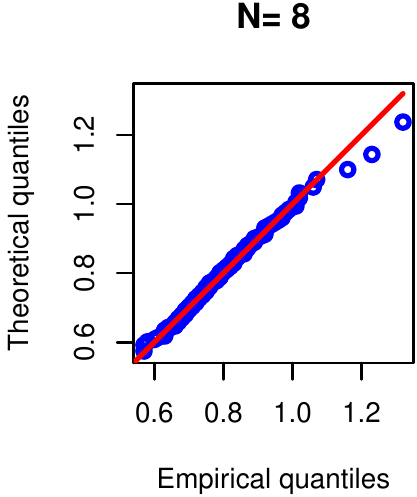}&
\includegraphics[width=1.05in]{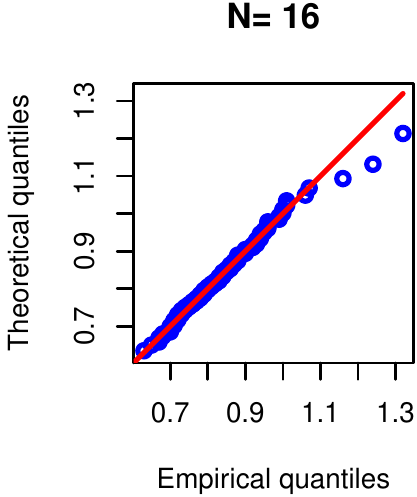}&
\includegraphics[width=1.05in]{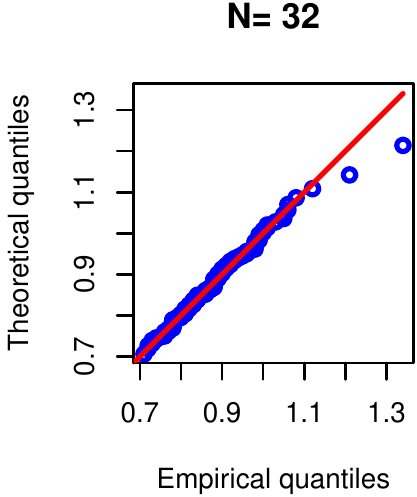}&
\includegraphics[width=1.05in]{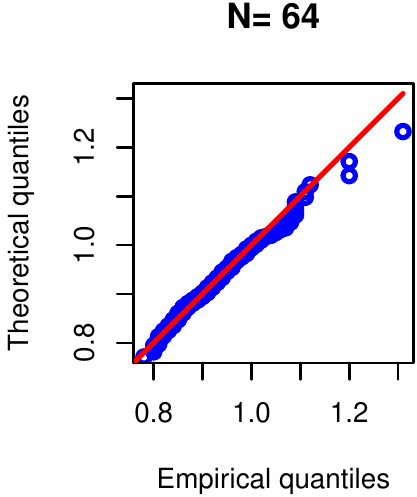}
\end{array}
$
\caption{\label{fig:qq-plot-3D} Quantile-quantile plot of empirical distribution of $\Lambda$ versus the fitted Gumbel distribution for 3D problems of sizes $N\in\{8, 16, 32, 64\}$. In each panel, the red line indicates the $y=x$ identity line.}
\end{figure*}

We perform Monte Carlo simulations to sample $\Lambda$ a total of $200$ times for each dimension $d$ and lattice size $N$.
We sample $\Lambda$ for $N\times N$ lattices of sizes $N\in \{8,16,32,64,128,256,512,1024\}$ for $d=2$,
and  for $N\times N \times N$ lattices of sizes $N\in \{8,16,32,64\}$ for $d=3$.
Since the definition of $\Lambda$ in~\eqref{eq:lambdaNd>1} involves maximum of random variables, we first considered 
the three parameter Generalized extreme value distribution to fit the $200$ empirical thresholds.
We then tested the two parameter Gumbel distribution against it.
The $p$-values of likelihood ratio tests, shown in Table~\ref{table-LRtest}, support fitting a Gumbel$(\mu,\beta)$ distribution.  
In Figures~\ref{fig:qq-plot-2D} and \ref{fig:qq-plot-3D}, the quantile-quantile plots show the empirical distribution of $\Lambda$ against the fitted Gumbel distribution for 2D and 3D problems respectively.
These plots show a good match between the empirical distributions and the fitted Gumbel distributions.

\begin{figure*}[ht]
\centering
\includegraphics[height=5in, angle=-90]{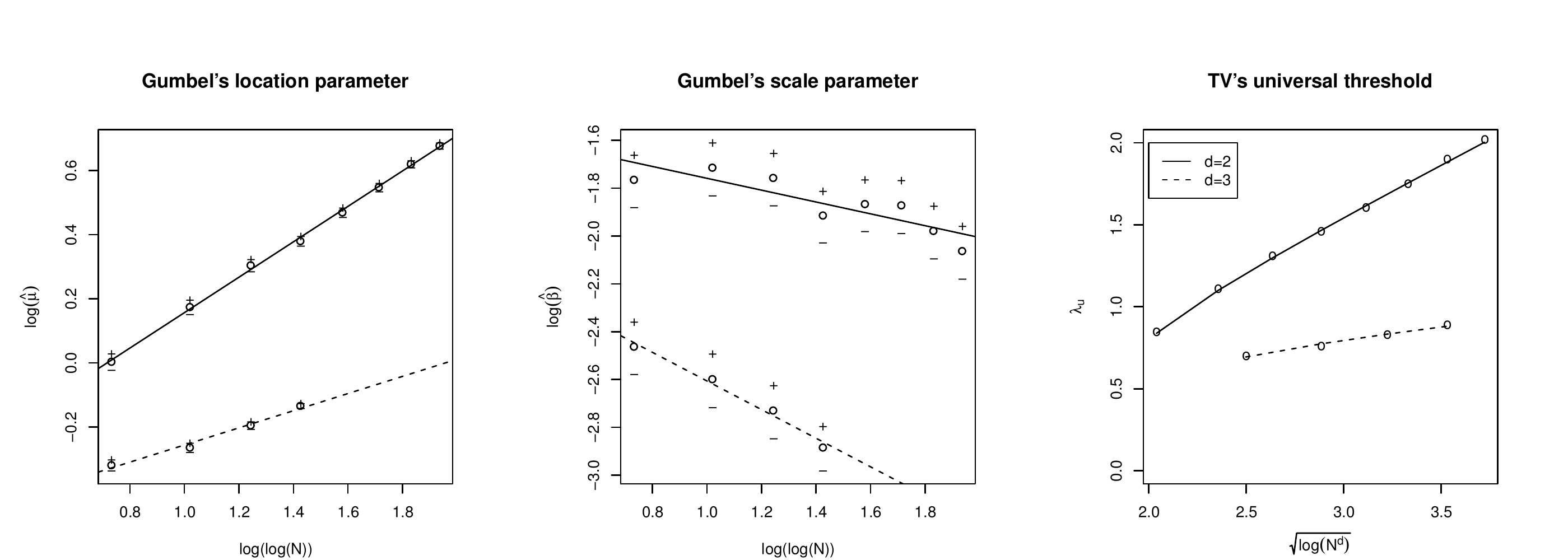}
\caption{\label{fig:Gumbel} Left and central plots: Gumbel$(\mu, \beta)$ fit of the sampled observations from the minimal threshold random variable $\Lambda$ defined in~\eqref{eq:lambdaNd>1} for $N^d$ square lattices for $d\in\{2,3\}$ and various $N$; Estimated parameter at $\circ$, plus $+$ or minus $-$ two standard errors; The lines are least squares fits. Right plot: estimated universal threshold as empirical quantiles (at $\circ$) and by Gumbel quantiles (lines).}
\end{figure*}

Based on the Gumbel approximation
$$
\Lambda_{N,d} \stackrel{\cdot}{\sim} {\rm Gumbel}(\mu(N,d), \beta(N,d)),
$$
we fit the following model
\begin{eqnarray}\label{eq:Gumbellinfit}
\nonumber {\mathbb E} \log \hat \mu(N,d)=a^\mu+b^\mu \log \log N,  
\\
{\mathbb E} \log \hat \beta(N,d)=a^\beta+b^\beta \log \log N,
\end{eqnarray}
for coefficients $a^\mu$, $b^\mu$, $a^\beta$, and  $b^\beta$. 
Figure~\ref{fig:Gumbel} shows the results of the fit. The left and middle panels show the estimated coefficients $\log \hat \mu(N,d)$  and 
$\log \hat \beta(N,d)$ as a function of $\log \log N$ for $d=2$ (continuous lines) and $d=3$ (dotted line).
Table~\ref{table-Gumbellinfit} provides the estimated values of
$(a^\mu, b^\mu, a^\beta, b^\beta)$ of the linear models.
The TV universal threshold shown on the right plots of Figure~\ref{fig:Gumbel} are then obtained by taking the $1-2/\sqrt{\log P_M}$ quantiles of the Gumbel$(\exp(\hat a^\mu+\hat b^\mu \log \log N), \exp(\hat a^\beta+\hat b^\beta \log \log N))$ distribution,
where $P_M=d N^{d-1}(N-1)$ is the number of finite differences involved in the TV penalty and $M=N^d$ is the problem size,
as discussed in Section~\ref{sct:dual}.

Algorithm~\ref{alg:adaTVimage} shows how the Gumbel approximation of the distribution of $\Lambda$ is used to denoise signals on a $d$-dimensional lattice with the adaptive universal threshold
that is an upper quantile of the Gumbel distribution for a given problem size $N^d$.
\begin{algorithm}
\caption{\label{alg:adaTVimage} TV denoising on a $d$-dimensional lattice with the adaptive universal threshold}
\label{TValgorithm}
\begin{algorithmic}[0]
\Procedure{$\hat f_2=\text{adaptiveTV}(y)$}{}
\State 1. Compute $\mu_1=\exp(\hat a^\mu+\hat b^\mu \log \log N)$ and $\beta_1=\exp(\hat a^\beta+\hat b^\beta \log \log N)$ based on Table~\ref{table-Gumbellinfit}.
\State 2. Calculate universal threshold  $\lambda_1 = \hat \sigma F_{\mu_1,\beta_1}^{-1}(1-2/\sqrt{\log P_M})$ with $P_M=dN^{d-1}(N-1)$ and
$\hat \sigma$ given in~\eqref{eq:MADsigma}. Here $F^{-1}_{\mu, \beta}$ is the inverse Gumbel, namely $F^{-1}_{\mu,\beta}(p)=\mu-\beta\log(-\log p)$.
\State 3. Calculate $\hat f_1$ by solving (\ref{eq:TVprimal}) using $\lambda_1$.
\State 4. Find significant number of connected components ${\rm NCC}(\hat \bbf_{\lambda_1})$  and compute average number $\bar N$ of observations per constant piece: $\bar{N}^d = N^d/{\rm NCC}(\hat \bbf_{\lambda_1})$.
\State 5. Repeat steps 1, 2 and 3 only once with $\bar N$ to get $\mu_2$, $\beta_2$, the adaptive universal threshold $\lambda_2$, and calculate $\hat f_2$.
\EndProcedure
\end{algorithmic}
\end{algorithm}


\begin{table}[ht]
\centering
\caption{Estimated coefficients of the linear fit \eqref{eq:Gumbellinfit} between the estimated Gumbel parameters $\hat \mu(N,d)$ and $\hat \beta(N,d)$ and $\log \log N$ for $d\in\{2,3\}$} 
\label{table-Gumbellinfit}
\begin{tabular}{|c|cccc|}
  \hline
  dimension& $\hat a^{\mu}$ & $\hat b^{\mu}$ & $\hat a^{\beta}$ & $\hat b^{\beta}$ \\ \hline
2 & -0.395 & 0.552 & -1.512 & -0.247 \\ 
  3 & -0.523 & 0.267 & -2.008 & -0.598 \\ 
   \hline
\end{tabular}
\end{table}

\subsection{Image denoising application} 
\label{subsct:imageex}

%

\begin{table*}[ht]
\tiny
\centering
\caption{Comparison of $\ell_2$ loss between the true image and the estimated image for oracle, SURE and adaptive universal threshold. The numbers reported are the percentage increase in $\ell_2$ loss with respect to the oracle choice of $\lambda$. Three noise levels: Low ($\sigma=5\sigma_f$), Medium ($\sigma=\sigma_f$) and High ($\sigma=\sigma_f/5$), where $\sigma_f$ is the ``standard error'' of the image.} 
\label{table-ImageMSEratio1}
\begin{tabular}{|r||ccc|ccc|ccc|ccc|ccc|ccc|}
  \hline
  & L & M & H & L & M & H & L & M & H & L & M & H & L & M & H & L & M & H \\ & \multicolumn{3}{c|}{barbara} & \multicolumn{3}{c|}{boat} & \multicolumn{3}{c|}{hill} & \multicolumn{3}{c|}{cameraman} & \multicolumn{3}{c|}{house} & \multicolumn{3}{c|}{pirate} \\ \hline
\underline{$\sigma$ known} &  &  &  &  &  &  &  &  &  &  &  &  &  &  &  &  &  &  \\ 
  SURE & 4 & 1 & 0 & 3 & 3 & 0 & 5 & 4 & 0 & 0 & 0 & 0 & 15 & 1 & 0 & 4 & 1 & 0 \\ 
  Adaptive & 1 & 0 & 15 & 4 & 1 & 3 & 5 & 4 & 2 & 9 & 0 & 2 & 23 & 2 & 0 & 1 & 0 & 3 \\ 
  \underline{$\hat \sigma$ with \eqref{eq:MADsigma}} &  &  &  &  &  &  &  &  &  &  &  &  &  &  &  &  &  &  \\ 
  SURE & 4 & 2 & 25 & 3 & 3 & 13 & 5 & 4 & 10 & 0 & 0 & 12 & 15 & 1 & 8 & 4 & 1 & 12 \\ 
  Adaptive & 1 & 0 & 46 & 4 & 2 & 13 & 5 & 4 & 10 & 8 & 0 & 7 & 23 & 2 & 1 & 1 & 0 & 11 \\ 
   \hline
\end{tabular}
\end{table*}
 
\begin{table*}[ht]
\tiny
\centering
\caption{See Table~\ref{table-ImageMSEratio1}.} 
\label{table-ImageMSEratio2}
\begin{tabular}{|r||ccc|ccc|ccc|ccc|ccc|ccc|}
  \hline
  & \multicolumn{3}{c|}{pentagon} & \multicolumn{3}{c|}{montage} & \multicolumn{3}{c|}{lena256} & \multicolumn{3}{c|}{lena512} & \multicolumn{3}{c|}{couple} & \multicolumn{3}{c|}{chess} \\ \hline
\underline{$\sigma$ known} &  &  &  &  &  &  &  &  &  &  &  &  &  &  &  &  &  &  \\ 
  SURE & 0 & 0 & 0 & 23 & 2 & 0 & 10 & 1 & 0 & 6 & 0 & 0 & 3 & 0 & 0 & 10 & 6 & 0 \\ 
  Adaptive & 15 & 2 & 21 & 14 & 0 & 2 & 29 & 9 & 5 & 1 & 1 & 0 & 5 & 4 & 4 & 8 & 19 & 13 \\ 
  \underline{$\hat \sigma$ with \eqref{eq:MADsigma}} &  &  &  &  &  &  &  &  &  &  &  &  &  &  &  &  &  &  \\ 
  SURE & 0 & 9 & 59 & 23 & 2 & 0 & 10 & 0 & 13 & 6 & 0 & 2 & 3 & 0 & 18 & 10 & 6 & 0 \\ 
  Adaptive & 15 & 1 & 74 & 14 & 0 & 0 & 28 & 7 & 18 & 1 & 0 & 1 & 5 & 5 & 17 & 8 & 19 & 12 \\ 
   \hline
\end{tabular}
\end{table*}

%
%
%

We have tested our two-step threshold selection on a collection of $12$ images ($d=2$) listed in Table~\ref{table-1DSegmentation}.
For each one, we simulated one noisy realization at three signal to noise ratios (SNR) defined as:
\emph{low} when $\sigma=5\sigma_f$, \emph{medium} when $\sigma=\sigma_f$, and \emph{high} when $\sigma=\sigma_f/5$, where $\sigma_f$ is the ``standard error'' of the image.
For each of the $36$ images, we applied TV denoising and calculated the $\ell_2$-loss between the true image and the TV estimate selecting $\lambda$ with five different methods:
\begin{itemize}
 \item oracle when the optimal $\lambda$ minimizes the $\ell_2$-loss between the TV estimate and the true image (unknown in practice).
 \item SURE when $\lambda$ minimizes an estimate of the $\ell_2$-risk with true $\sigma$.
 \item SURE when $\lambda$ minimizes an estimate of the $\ell_2$-risk with $\sigma$ estimated by \eqref{eq:MADsigma}.
 \item universal threshold defined by $\lambda_2$ in algorithm~1 with true $\sigma$.
 \item universal threshold defined by $\lambda_2$ in algorithm~1 with $\sigma$ estimated by \eqref{eq:MADsigma}.
\end{itemize}
The goal is to compare threshold selection methods for TV. The oracle selection represents an unachievable benchmark because it requires knowing the true image.
Table~\ref{table-1DSegmentation} reports the percentage of increase in the $\ell_2$-loss with respect to the oracle choice of $\lambda$.
We observe that the adaptive universal threshold performs remarkably well in comparison with the others.

\begin{figure*}[ht]
\centering
\includegraphics[width=4.7in,angle=0]{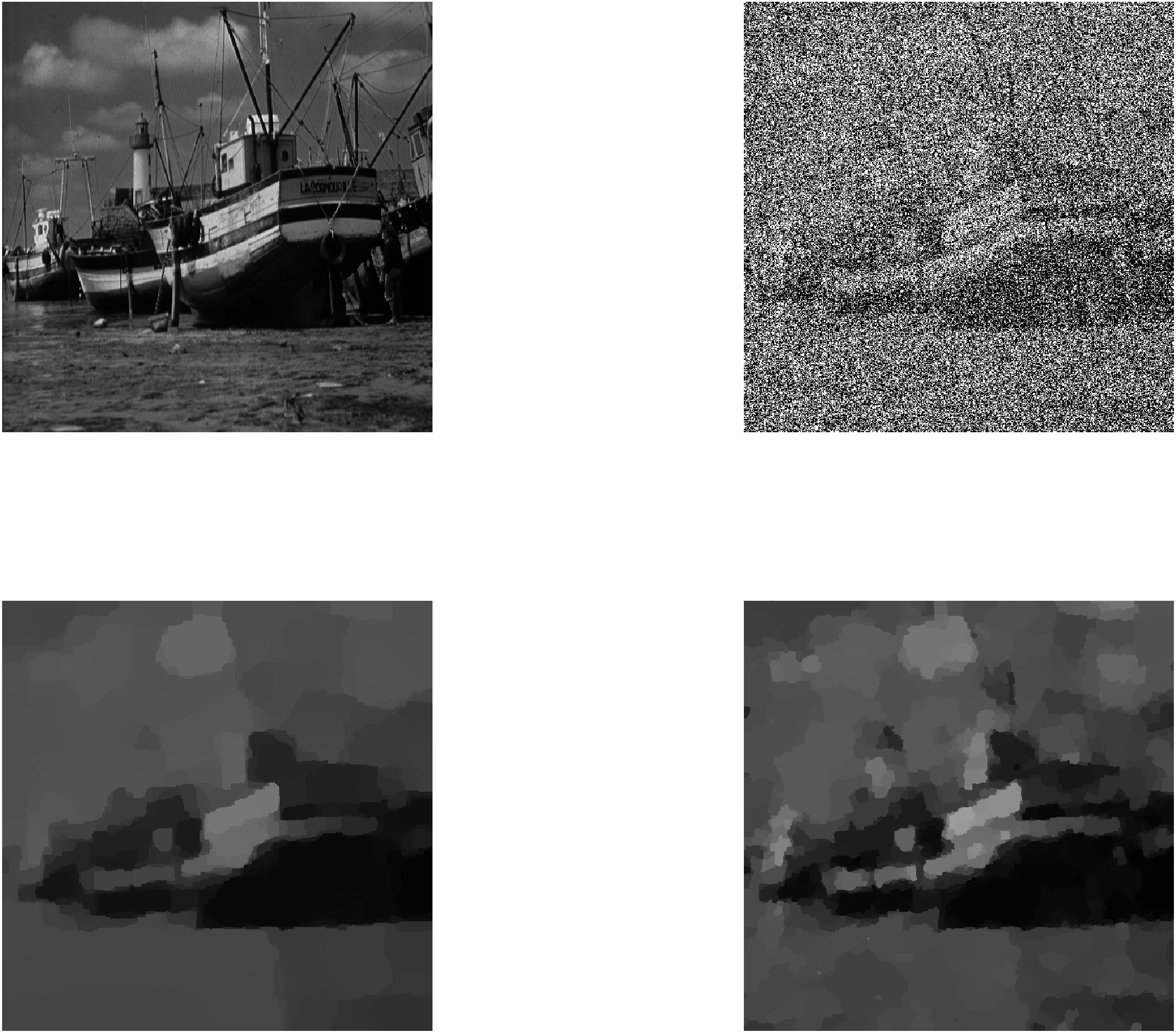}
\caption{Result for "boat" image with low SNR ($\sigma=5\sigma_f$). Top left: true image. Top right: noisy image. Bottom left: First step TV estimate with universal threshold.
Bottom right: second step TV estimate with adaptive threshold.}
\label{fig:image1}
\end{figure*}

\begin{figure*}[ht]
\centering
\includegraphics[width=4.7in,angle=0]{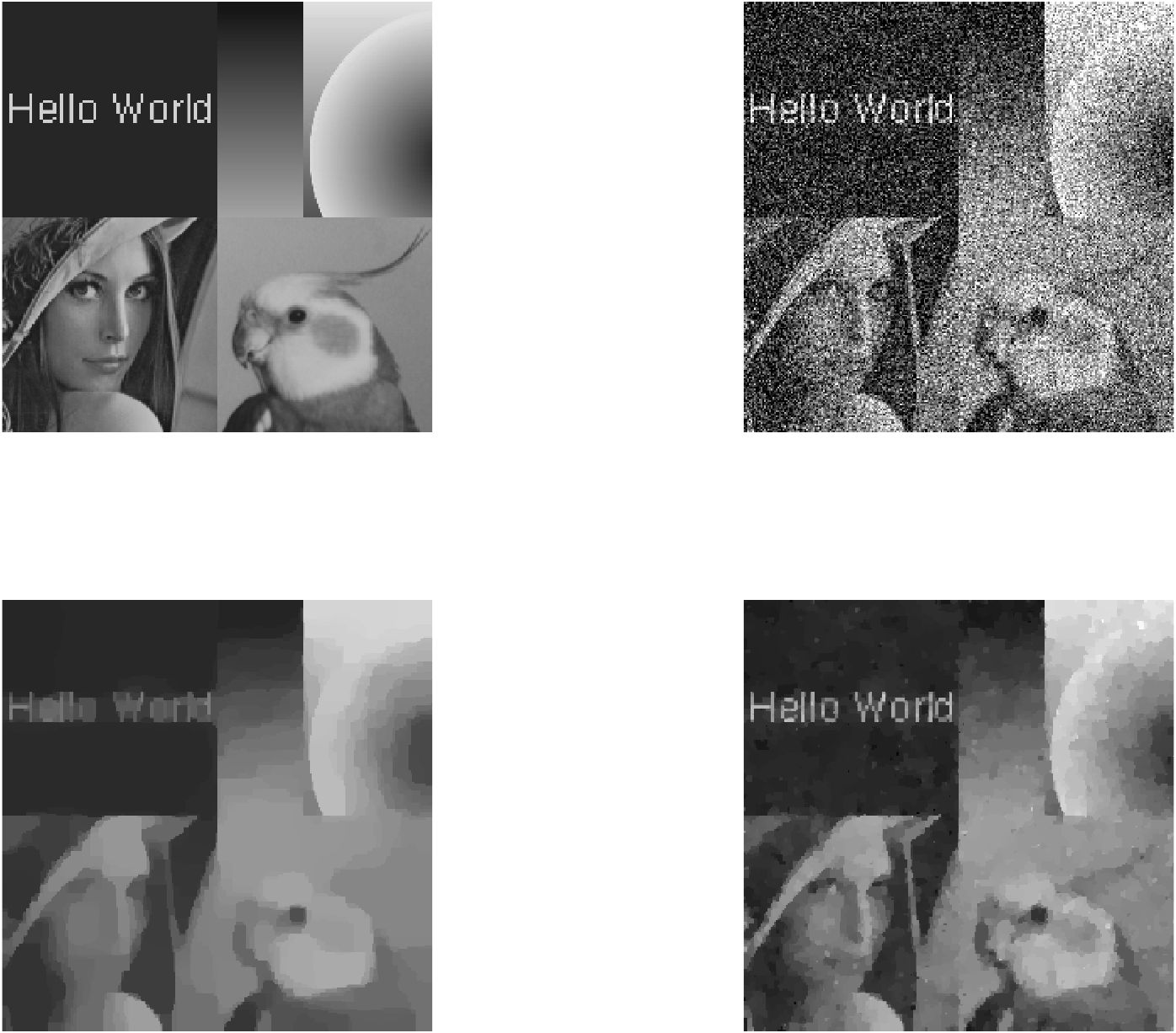}
\caption{Result for ``montage" image with medium SNR ($\sigma=\sigma_f$). Top left: true image. Top right: noisy image. Bottom left: First step TV estimate with universal threshold.
Bottom right: second step TV estimate with adaptive threshold.}
\label{fig:image2}
\end{figure*}

\begin{figure*}[ht]
\centering
\includegraphics[width=4.7in,angle=0]{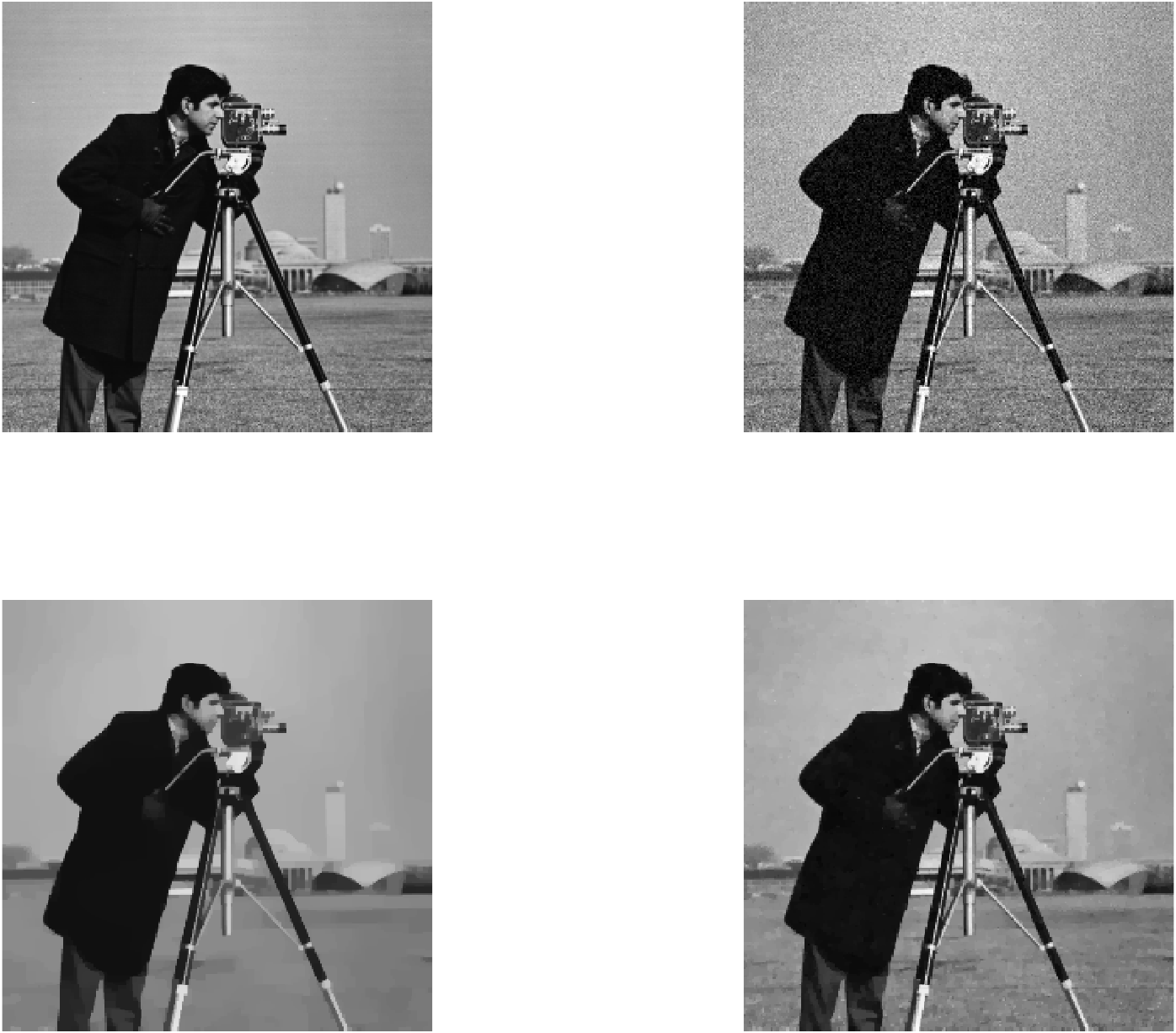}
\caption{Result for ``cameraman" image with high SNR ($\sigma=\sigma_f/5$). Top left: true image. Top right: noisy image. Bottom left: First step TV estimate with universal threshold.
Bottom right: second step TV estimate with adaptive threshold.}
\label{fig:image3}
\end{figure*}

\begin{figure*}[ht]
\centering
$
\begin{array}{lll}
\includegraphics[width=1.5in]{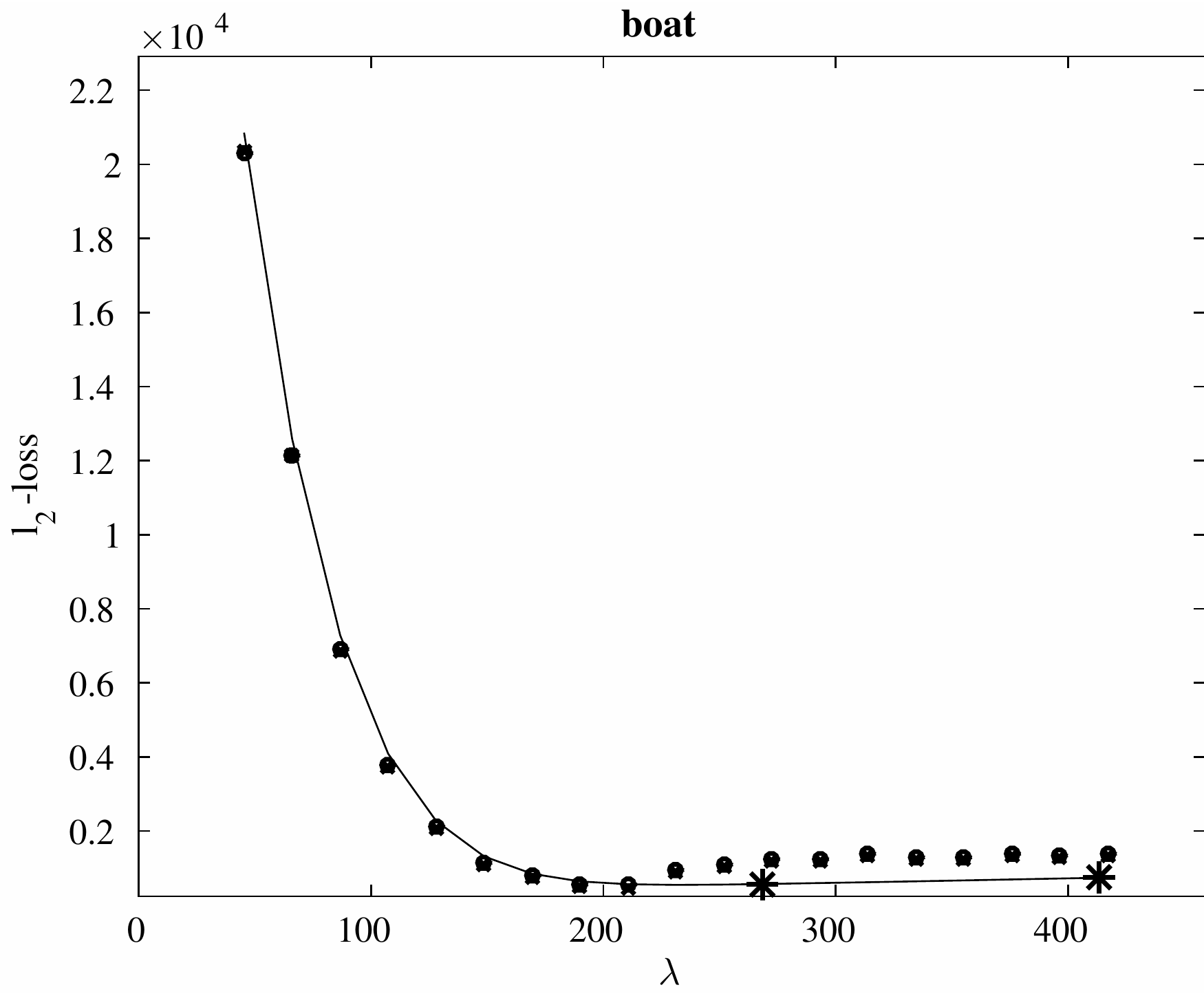}&
\includegraphics[width=1.5in]{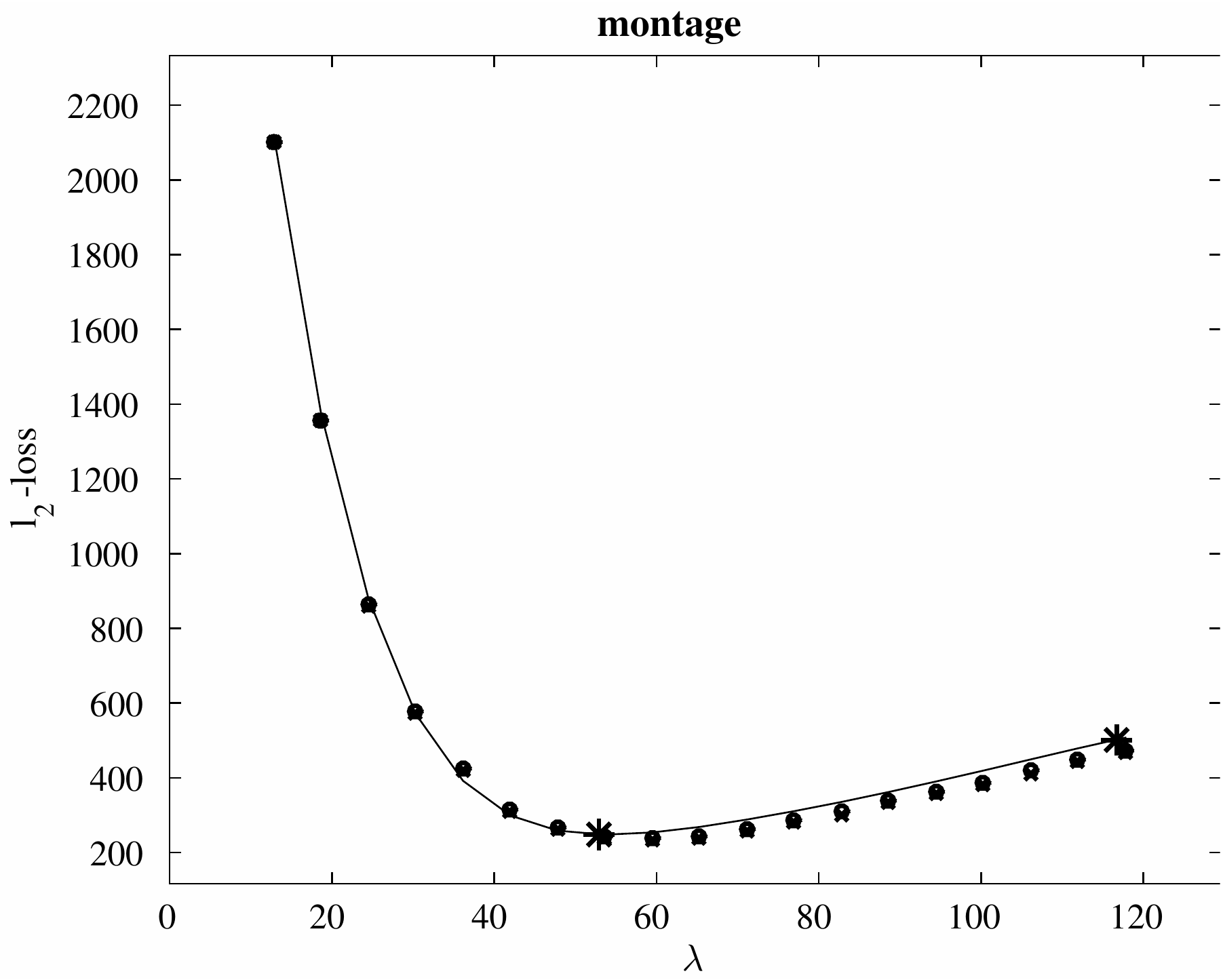}&
\includegraphics[width=1.5in]{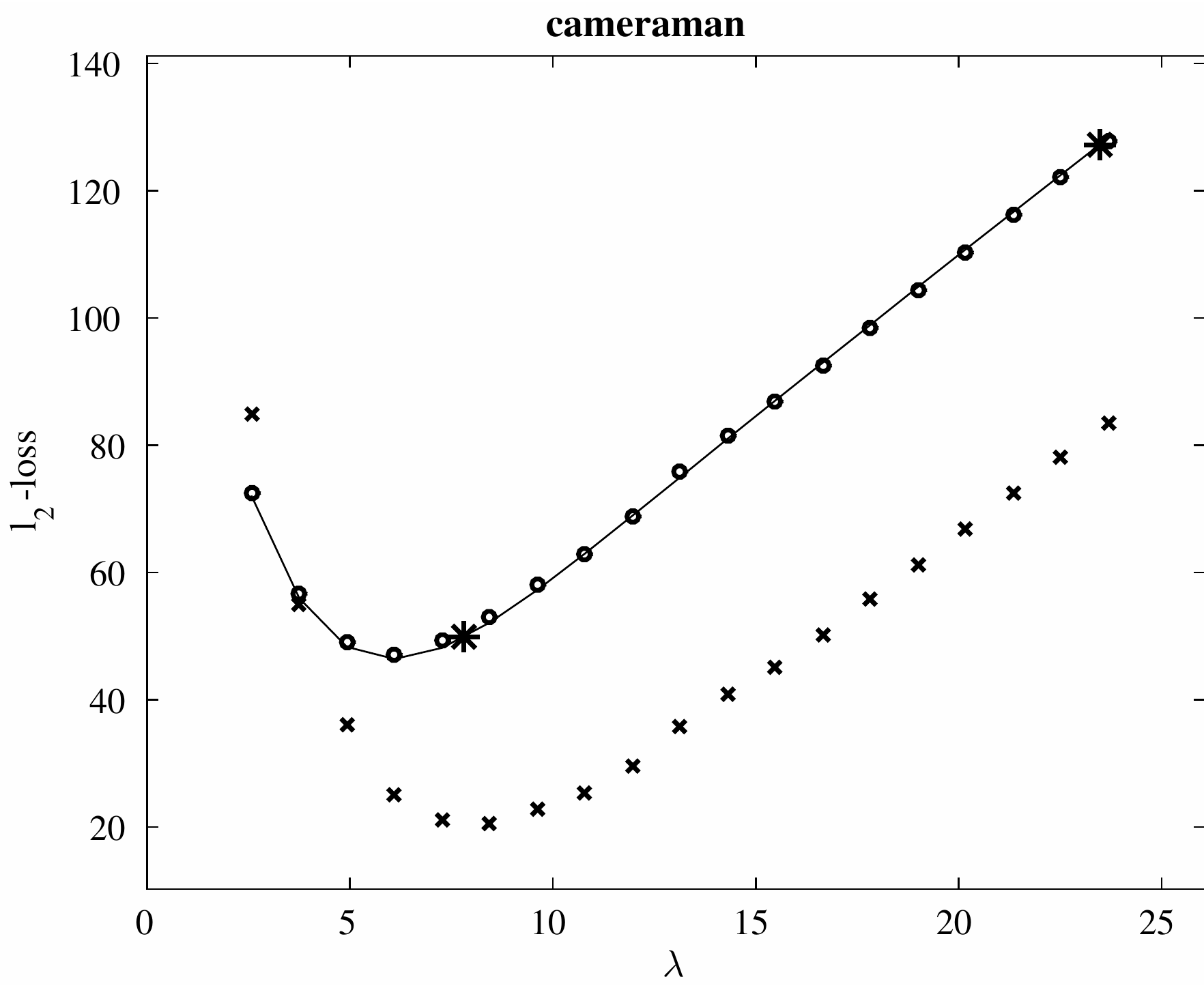}
\end{array}
$
\caption{Quality of image reconstruction as a function of $\lambda$ corresponding to Figures~\ref{fig:image1}, \ref{fig:image2} and \ref{fig:image3}.
True $\ell_2$-loss (line); Stein unbiased risk estimate with $\sigma$ known (o) and estimated with~\eqref{eq:MADsigma} (x);
First step (right *) and second step (left *) TV estimate.}
\label{fig:l2loss}
\end{figure*}

To further illustrate, we have plotted three of these images in Figures~\ref{fig:image1}--\ref{fig:image3},
whose corresponding mean squared errors are plotted in Figure~\ref{fig:l2loss}. 

In all cases the two-step procedure gives a selection of the threshold close to the optimal value, while SURE$(\lambda)$ needs to track down a minimum by trying many $\lambda$.
For more examples and results, see Section~\ref{rr}.

\section{Conclusions}
\label{sct:conclusion}

We have presented an efficient procedure for selecting the threshold in total variation denoising.
Our methodology is 
\emph{adaptive} in the sense that our threshold adapts to the complexity of the underlying function as measured by its number of connected components.
We applied our method to denoise various 1D and 2D signals, and we observed remarkable performance in terms of mean-squared errors.
We also studied the ability of total variation to perform exact segmentation for 1D signals. While the empirical processes involved in the selection of $\lambda$ 
in the 1D case are well studied (Brownian bridge with drifts), the empirical processes encountered in the 2D and 3D cases appear to be mathematically challenging.
The empirical data suggest a logarithmic rate of growth for these processes, but a rigorous justification seems to be beyond the currently known results in probability theory. 

\section{Reproducible Research} \label{rr}

The code and data that generated the figures in this article may be found online
at {\tt http://purl.stanford.edu/sw114yc8625}\cite{TSTVD-PURL}.


\section{Acknowledgements}

We are grateful to Jon A. Wellner for his help on a proof regarding the large deviation probabilities of certain empirical processes encountered in this work.
We would also like to thank Michael Saunders and Jairo Diaz Rodriguez for their helpful comments and discussions, Peyman Milanfar for providing test images, and the Stanford Research Computing Center for providing computational resources and support that have contributed to these research results.
This work was funded by the Swiss National Science Foundation and by the National Science Foundation Grants DMS0906812  (American Reinvestment and Recovery Act) and
DMS-1418362 (`Big-Data' Asymptotics: Theory and Large-Scale Experiments).

\appendix

\section{Proof of Theorem~3.1}
\begin{proof}
Total variation can be rewritten as a lasso estimator in dimension one.
Let ${\boldsymbol \mu}={\bf 1}\alpha_0 + X {\boldsymbol \alpha}$, where $[{\bf 1},X]$ is the (invertible) lower triangular matrix of ones.
With this change of variable, the finite differences are $\alpha_i=f_{n+1}-f_n$ for $n=1,\ldots,N-1$ and $\alpha_0=f_1$,
and one can rewrite TV as solution to
\begin{equation} \label{eq:tv=lasso}
\min_{\alpha_0,{\boldsymbol \alpha}} \frac{1}{2} \| {\bf y}-\alpha_0 {\bf 1} - X {\boldsymbol \alpha} \|_2^2 + \lambda \| {\boldsymbol \alpha} \|_1.
\end{equation}

The dual of lasso has been derived \citep{Osbo:Pres:Turl:on:2000}. So consider the lasso formulation of total variation~(\ref{eq:tv=lasso})
and let $X^*$ be the oracle columns of $X$ (those corresponding to a jump), let  $X_-^*$ be the non-oracle columns of $X$,
let ${\bf s}$ be the sign of the oracle jumps, and let  $P=H/N$ be the projection matrix for the intercept on the column vector of ones, where $H$ is the $N \times N$ matrix of ones.
The oracle KKT conditions are:
$$
\left \{
\begin{array}{rcl}
\alpha_0&=& \frac{1}{N}{\bf 1}^{\rm T} ({\bf y}-X^* {\boldsymbol \alpha}^*) \\ 
(X^*)^{\rm T} (I-P)({\bf y}-X^*{\boldsymbol \alpha}^*)&=& \lambda {\bf s} \\
{\bf s} &=& {\rm sign} ({\boldsymbol \alpha}^*) \\
(X^*_-)^{\rm T} (I-P) ({\bf y}-X^* {\boldsymbol \alpha}^*) &\in& [-\lambda,\lambda]
\end{array}
\right . 
$$
$$
\Longleftrightarrow \left \{
\begin{array}{rcl}
\alpha_0&=& \frac{1}{N}{\bf 1}^{\rm T} ({\bf y}-X^* {\boldsymbol \alpha}^*) \\ 
\hat {\bbf} &=& \alpha_0 {\bf 1} + X^* {\boldsymbol \alpha}^* \\
(X^*)^{\rm T} ({\bf y}-\hat {\bbf})&=& \lambda {\bf s} \\
{\bf s} &=& {\rm sign} ({\boldsymbol \alpha}^*) \\
(X^*_-)^{\rm T} ({\bf y}-\hat {\bbf}) &\in& [-\lambda,\lambda].
\end{array}
\right .
$$
Since the matrix $X$ has columns representing Heaviside functions at the oracle jump locations, its transpose $X^{\rm T}$ amounts to performing partial sums; likewise for $(X^*)^{\rm T}$ and $(X_-^*)^{\rm T}$. 
Another consequence is that the estimated function is the vector $\hat {\bbf}=(\hat {\bbf}_1, \ldots, \hat {\bbf}_L)$, where each $\hat {\bbf}_1=\hat h_l {\bf 1}_{N_l}$ is a constant vector of length $N_l$ for $l=1,\ldots,L$.
Likewise, ${\rm sign}({\boldsymbol \alpha}^*)={\rm sign}(\hat h_{l+1}-\hat h_l)$.
After solving in $\hat {\bbf}$ the linear system $(X^*)^{\rm T} ({\bf y}-\hat {\bbf})= \lambda {\bf s}$ of $L$ equations with essentially $L$ unknowns because $\hat {\bbf}$ is a vector with $L$ constant pieces,
the KKT conditions can equivalently be written as
\begin{equation*} 
\left \{
\begin{array}{rcl}
\hat {\bbf}_l&=&\hat h_l {\bf 1}_{N_l}, \quad l=1,\ldots,L \\
\hat h_l &=& \bar y_j + (s_l-s_{l-1}) \frac{\lambda}{N_l}, \quad l=1,\ldots,L \\
 {\rm sign}(\hat h_{l+1}-\hat h_{l})&=&s_l, \quad l=1,\ldots,L-1\\
\bw &\in& [-\lambda,\lambda]
\end{array}
\right . ,
\end{equation*}
where the dual vector $\bw=(X^*_-)^{\rm T} ({\bf y}-\hat {\bbf})$.
%
After some algebra, the dual vector is
$$
w_{N_{\bullet l-1}+i}= -\{ \sum_{k=1}^i  y_{N_{\bullet l-1}+k} -i \bar y_j  \}  + \frac{\lambda}{N_l} i (s_{l}-s_{l-1}) + \lambda s_{l-1},
$$
for $i=1,\ldots,N_l$ and $l=1,\ldots,L$.

\section{Proof of Theorem~3.2}

Observing that $( \sum_{k=1}^i y_{N_{\bullet l-1}+k}-i \bar y_j )/\sqrt{N_l}$ for $i=1,\ldots,N_l$
has the distribution of a discretized Brownian bridge on a grid $t_i=i/N_l, \ i=1,\ldots,N_l$, we know that the dual vector $\bw=(\bw_1 ,\ldots,\bw_L)$ with
$\bw_j=(w_{N_{\bullet l-1}+1},\ldots,w_{N_{\bullet l-1}+N_l})$ behaves asymptotically like independent Brownian bridges $\mathbb{U}_j$
with a drift, namely
$$
\bw_{l,i} = \sqrt{N_l} \mathbb{U}_j(t_i) + \lambda  t_i (s_j-s_{l-1}) + \lambda s_{l-1}.
$$
And a necessary condition for the KKT conditions to be satisfied is 
$$
\{\| \bw_j \|_\infty \leq \lambda\}
$$
for all $l=1,\ldots,L$. We seek $\lambda$ such that, for all $l=1,\ldots,L$ (where $L$ is fixed), we have $\prob (\| \bw_j \|_\infty \leq \lambda)\rightarrow 1$ as $N$ (hence $N_l$) goes to infinity. 
The number of jumps $L$ being fixed and the discretized Brownian bridges being independent, we can simply control the probability going to one for each individual $l=1,\ldots,L$.
We first have 
\begin{eqnarray}
\prob (\| \bw_j \|_\infty \leq \lambda) &\geq& \prob(-\lambda \leq \sqrt{N_l} \mathbb{U}_j(t)  \nonumber \\
&& +\lambda  t (s_j-s_{l-1}) + \lambda s_{l-1} \leq \lambda, \nonumber \\
&& \mbox{for all} \ t \in [1/N_l,1-1/N_l]) \nonumber \\
&\geq& \prob(-\lambda \leq \sqrt{N_{\max}} \mathbb{U}_j(t) \nonumber \\ 
&& + \lambda  t (s_j-s_{l-1}) + \lambda s_{l-1} \leq \lambda, \nonumber \\
&&  \mbox{for all} \ t \in [1/N_{\max},1-1/N_{\max}]), \label{eq:lower1}
\end{eqnarray}
where $N_{\max}=\max_{l=1,\ldots,L} N_l$.
Three cases must be considered for the lower probability in~(\ref{eq:lower1}):
\begin{enumerate}
 \item boundary case, e.g., $s_{L-1}=-1$ and $s_L=0$: 
\begin{eqnarray*}
  &&\prob(-\lambda t \leq \sqrt{N_{\max}} \mathbb{U}_j(t)  \leq \lambda (2-t), \\ &&  \quad \mbox{for all} \ t \in [1/N_{\max},1-1/N_{\max}])
\end{eqnarray*}
 \item interior case with sign change, e.g., $s_l=-1=-s_{l+1}$:
\begin{eqnarray*}
 && \prob(-2\lambda t \leq \sqrt{N_{\max}} \mathbb{U}_j(t)  \leq 2 \lambda (1-t), \\ && \quad  \mbox{for all} \ t \in [1/N_{\max},1-1/N_{\max}])
\end{eqnarray*}
 \item interior case with no sign change, e.g., $s_j=-1=s_{l+1}$:
\begin{eqnarray*}
 && \prob(0 \leq \sqrt{N_{\max}} \mathbb{U}_j(t)  \leq 2 \lambda, \\ && \quad  \mbox{for all} \ t \in [1/N_{\max},1-1/N_{\max}]).
\end{eqnarray*}
\end{enumerate}
Clearly case 3 leads to a probability tending to zero as $N_{\max}$ goes to infinity because the probability of a Brownian bridge to be always positive on $[0,1]$ is essentially zero.

So we concentrate on the case where the jump signs alternate. The two boundaries drive the order of the convergence rate, so let us consider case~1 first:
\begin{eqnarray*}
 \prob(-\lambda t \leq \sqrt{N_{\max}} \mathbb{U}(t) \leq \lambda (2-t))&\geq &1-\prob_{\lambda}(A )-\prob_{\lambda}(B ),
\end{eqnarray*}
where:
\begin{itemize}
 \item $A=\{\sqrt{N_{\max}} \mathbb{U}(t)< -\lambda t, \ \mbox{for some}\ t \in [1/N_{\max},1-1/N_{\max}]\}$,
 \item $B=\{\sqrt{N_{\max}}\mathbb{U}(t)> \lambda(2-t), \ \mbox{for some}\ t \in [1/N_{\max},1-1/N_{\max}]\}$.
\end{itemize}
 We have
\begin{eqnarray*}
 \prob_{\lambda}(A )&\approx& \prob(\sqrt{N_{\max}} \mathbb{U}(t)< -\lambda t, \ \mbox{for some}\ t \in [1/N_{\max},1])\\
&=& 2 (1-\Phi(\frac{\lambda}{\sqrt{N_{\max}}}\sqrt{\frac{1/N_{\max}}{1-1/N_{\max}}}))\\
& \approx& 2 (1-\Phi(\frac{\lambda}{N_{\max}}))
\end{eqnarray*}
by (24) of page 39 of \citet{Shor:Well:empi:1986}.
For $\prob_{\lambda}(A)=\alpha_N$ for $\alpha_N$ small, we want 
\begin{equation} \label{eq:uniTV0}
\lambda_N= N_{\max}\Phi^{-1}(1-\alpha_N/2).
 \end{equation}
Moreover,
\begin{eqnarray*}
 \prob_{\lambda}(B) &=& \prob_{\lambda}(\sqrt{N_{\max}}\mathbb{U}(t)> \lambda (2-t),  \\&& \quad \mbox{for some}\ t \in [1/N_{\max},1-1/N_{\max}])\\
 &\leq& \prob_{\lambda}(\sqrt{N_{\max}} \mathbb{U}(t)> \lambda (1+1/N_{\max}), \\ && \quad \mbox{for some}\ t \in [0,1])\\
 &=&\exp(-2\lambda^2/N_{\max}(1+1/N_{\max})^2;
\end{eqnarray*}
so, for $\lambda=\lambda_N$ in~(\ref{eq:uniTV0}), this latter probability is exponentially small in $N$ even if $\alpha_N$ is constant.
And for case~2 (the interior jumps), we have
\begin{eqnarray*}
 \prob(-2\lambda t \leq \sqrt{N_{\max}} \mathbb{U}(t) \leq 2 \lambda (1-t))&\geq &1-2 \prob_{2\lambda}(A ).
\end{eqnarray*}
Putting the inequalities together for the $L-2$ interior jumps and  the two boundary jumps, we bound from below the probability
for TV to perform exact segmentation:
\begin{eqnarray} \label{eq:pi0ES}
\prob (\max_{l=1,\ldots,L}\| \bw_j \|_\infty \leq \lambda_N) &\geq& (1-2\alpha_N)^{L-2} (1-\alpha_N)^2 \nonumber \\ &=:&\pi_0^{\rm ES}. 
\end{eqnarray}


%
%
\end{proof}

\bibliographystyle{plainnat}
\bibliography{article}


\end{document}